\documentclass{aa}  

\usepackage{upgreek}
\usepackage{txfonts}
\usepackage{graphicx}
\usepackage{upgreek}
\usepackage{xcolor}
\usepackage{algpseudocode}
\usepackage{algorithm}
\usepackage{txfonts}
\DeclareMathOperator*{\argmin}{arg\,min}

\newcommand{\prop}[1]{C\left\{#1\right\}}

\newcommand{\bigo}[1]{\mathcal{O}\left(#1\right)}
\newcommand{\edited}[1]{\textcolor{black}{#1}}
\newcommand{\editedLC}[1]{\textcolor{black}{#1}}
\begin{document}

   \title{Implicit electric field Conjugation: Data-driven focal plane control}

   \author{S.~Y.~Haffert\inst{1}\fnmsep\thanks{NASA Hubble Fellow}
   \and 
   J.~R.~Males\inst{1}
   \and
   K. Ahn\inst{2}
    \and 
   K.~Van Gorkom\inst{1}
   \and
   O. Guyon\inst{1,2,3,4}
   \and
   L.~M.~Close\inst{1}
   \and
   J. D. Long\inst{1}
   \and
   A. D. Hedglen\inst{1,3}
   \and
   L. Schatz\inst{5}
   \and
   M. Kautz\inst{3}
   \and
   J. Lumbres\inst{3}
   \and 
   A. Rodack\inst{3}
   \and
   J. M. Knight\inst{3}
   \and
   K. Miller\inst{6}
   }

   \institute{University of Arizona, Steward Observatory, Tucson, Arizona, United States
   \and
    National Astronomical Observatory of Japan, Subaru Telescope, National Institutes of Natural Sciences, Hilo, HI 96720, USA
    \and
    Wyant College of Optical Science, University of Arizona, 1630 E University Blvd, Tucson, AZ 85719, United States
    \and
    Astrobiology Center, National Institutes of Natural Sciences, 2-21-1 Osawa, Mitaka, Tokyo, JAPAN
    \and
    Air Force Research Laboratory, Directed Energy Directorate, Space Electro-Optics Division, Starfire Optical Range, Kirtland Air Force Base, New Mexico, United States
    \and
    LWS Optics and Beam Control Branch, Naval Surface Warfare Center, Virginia, United States \\
    \email{shaffert@arizona.edu}
   }
   \date{Received 12 September 2022 / Accepted 5 February 2023}

% \abstract{}{}{}{}{} 
% 5 {} token are mandatory
 
 %
 %
 
  \abstract
  % context heading (optional)
  % {} leave it empty if necessary  
   {Direct imaging of Earth-like planets is one of the main science cases for the next generation of extremely large telescopes. This is very challenging due to the star-planet contrast that has to be overcome. Most current high-contrast imaging instruments are limited in sensitivity at small angular separations due to non-common path aberrations (NCPA). The NCPA leak through the coronagraph and create bright speckles that limit the on-sky contrast and therefore also the post-processed contrast.}
  % aims heading (mandatory)
   {We aim to remove the NCPA by active focal plane wavefront control using a data-driven approach.}
  % methods heading (mandatory)
   {We developed a new approach to dark hole creation and maintenance that does not require an instrument model. This new approach is called implicit Electric Field Conjugation (iEFC) and it can be empirically calibrated. This makes it robust for complex instruments where optical models might be difficult to realize. Numerical simulations have been used to explore the performance of iEFC for different coronagraphs. The method was validated on the internal source of the Magellan Adaptive Optics eXtreme (MagAO-X) instrument to demonstrate iEFC's performance on a real instrument.}
  % results heading (mandatory)
   {Numerical experiments demonstrate that iEFC can achieve deep contrast below $10^{-9}$ \edited{with} several coronagraphs. The method is easily extended to broadband measurements and the simulations show that a bandwidth up to 40\% can be handled without problems. Lab experiments with MagAO-X showed a contrast gain of a factor 10 in a broadband light and a factor 20 to 200 in narrowband light. A contrast of $5\cdot10^{-8}$ was achieved with the Phase Apodized Pupil Lyot Coronagraph \edited{at 7.5 $\lambda/D$}. }
  % conclusions heading (optional), leave it empty if necessary 
   {The new iEFC method has been demonstrated to work in numerical and lab experiments. It is a method that can be empirically calibrated and it can achieve deep contrast. This makes it a valuable approach for complex ground-based high-contrast imaging systems.}


   \maketitle
%
%________________________________________________________________

\section{Introduction}
The next generation of giant segmented mirror telescopes will have the unprecedented angular resolution and sensitivity to directly image Earth-like planets around other stars. With these extremely large telescopes, we may even be able to detect bio signatures in their atmospheres. However, ground-based telescopes face a challenge by imaging through Earth's atmosphere, which degrades the spatial resolution due to turbulence. High-contrast imaging (HCI) instruments are designed to overcome these challenges by using extreme adaptive optics (ExAO) to compensate for atmospheric disturbances and recover the angular resolution \citep{guyon2018exao}. After restoring the intrinsic spatial resolution of the instrument, advanced coronagraphs remove the diffraction pattern \edited{of the on-axis star to create dark regions, also called dark holes,} where planets can be observed.

%The current generation of HCI instrument \citep{macintosh2014gpi, jovanovic2015scexao, males2018magaox, beuzit2019sphere} routinely reaches post-processed contrast levels of $10^{-4}$ to $10^{-6}$, depending on the angular distance from the host star. These contrast levels are sufficiently sensitive to observe hot and massive self-luminous planets \citep{marley2007hotjupiters}. Even though we are sensitive to massive Jupiter-like planets, only several planets have been imaged and spectroscopically characterized. The results from large surveys that targeted massive Jupiters on wide orbits, indicate that the planet occurrence rate drops sharply between 1 to 10 AU \citep{bowler2015gpoccurence, nielsen2019gpies, fernandes2019occurence}. More planets could be directly imaged if the sensitivity close to the star is improved. 

The NCPA are wavefront errors that are not seen by the AO systems. \edited{The AO system which use beamsplitters to split the light into a part that is used for wavefront sensing and a part for the science instruments. The NCPA arise from errors that are created in the science optical path after the light is split off for wavefront sensing.} The NCPA are generated inside the instrument and evolve slowly in time due to temperature changes or changes in the gravity vector. The NCPA create a speckle halo that can mimic exoplanet signals and, due to their slow temporal evolution, they do not average out over typical observing timescales. Advanced image processing techniques are necessary to remove the stellar speckles and recover the planet signal. But post-processing algorithms that depend on spatial diversity, such as angular differential imaging (ADI) \citep{marois2006adi}, are not able to remove the speckles efficiently at small angles due to the limited spatial diversity. 

\editedLC{Active correction of the NCPA} during the observations would sidestep this issue. To achieve active compensation, we need to be able to sense the wavefront at the science focal plane, which is typically achieved with a Focal Plane Wavefront Sensor (FPWFS). Sensing the wavefront at any other plane would still allow for the existence of NCPA. Removing NCPA and maintaining an aberration-free system would drastically improve the sensitivity at small angular separations. Several FWPFS have been implemented on test benches and validated with or without residual atmospheric turbulence \citep{singh2019active, potier2019exoplanet, herscovici2019coronagraphic}. Few on-sky FPWFS control experiments have been done. One of the first was speckle nulling with \editedLC{the Subaru Coronagraphic Extreme Adaptive Optics instrument (SCExAO)} \citep{martinache2014sky} and the same technique was later applied on Palomar P1640 and at Keck \citep{bottom2016speckle}. SCExAO has demonstrated several other FPWFS for control, such as single-shot FPWFS with the vector-Apodizing Phase Plate (vAPP) coronagrah \cite{bos2019vappwfs}, Fast\&Furious \citep{wilby2018laboratory,bos2020sky}, and linear dark field control (LDFC) \cite{miller2017spatial, miller2021spatial}. All methods have resulted in moderate gain on-sky due to various reasons.

Speckle nulling has to take at least four images per speckle. Only a few speckles can be probed at the same time to minimize crosstalk. This makes speckle nulling a relatively slow algorithm and only very slowly changing quasi-statics can be removed. The single-shot vAPP FPWFS requires a lot of computational power, which limited the reconstruction to several tens of zernike modes. This enabled vAPP FPWFS to stabilize the PSF, but the speckles in the dark hole could not be removed. Fast\&Furious does run at high speed and can control high-order modes, but it can only work with a noncoronagraphic image. LDFC runs at high speeds (>100 Hz) and can converge to a (coronagraphic) reference \editedLC{Point Spread Function (PSF)}. However, LDFC can not make a dark hole by itself. This means that another algorithm has to create a coronagraphic dark hole first.

A standard approach to FPWFS is pair-wise probing (PWP) \citep{give2011pair}. In PWP, a phase probe is added to the deformable mirror (DM) to modulate the focal plane speckles. The full electric field can be reconstructed by stepping through several phase offsets of this probe. This is similar to the speckle nulling technique; however, the full region of interest is probed at once, instead of a single Fourier mode being probed at a time. A model of the instrument is used to reconstruct the electric field after all images are measured. This sensing approach is often combined with electric field conjugation (EFC) \citep{give2007broadband}. EFC tries to exactly cancel the electric in the focal plane by injecting equal strength speckles with an opposite phase. Similar to the PWP electric field reconstruction, the instrument model is used to derive the optimal shape of the DM. The combination of PWP with EFC has been used on vacuum test benches to reach $10^{-10}$ contrast \citep{serabyn2013high, ruane2022broadband}, and more recently it has been used to create a dark hole with the Spectro-Polarimetric High-contrast Exoplanet REsearch (SPHERE) instrument \citep{beuzit2019sphere} \edited{on the internal source \citep{potier2020efcsphere} and on-sky \citep{potier2022increasing}}.

EFC is inherently a model-dependent algorithm, which means that the final performance and convergence speed depend on the accuracy of the system model. Any model error leads to reduced performance.  We propose the implicit EFC (iEFC) method, which uses the linear relation between the DM response and modulated intensity measurements. This sidesteps the intermediate electric field reconstruction step, allowing us to create dark holes without having a system model. \edited{A similar approach has been proposed before \citep{ruffio2022non}.} The theoretical background of the PWP sensing strategy, EFC and iEFC are derived in Section 2. In Section 3, the performance of iEFC is demonstrated numerically for different aperture geometries and coronagraphs. Section 4 shows the experimental validation of iEFC by creating dark holes on several test beds with different \editedLC{coronagraphs} to demonstrate the ease and robustness of iEFC.
  
%__________________________________________________________________
\section{Theoretical background}
\subsection{The focal plane intensity}
The electric field of an aberrated wavefront that is modulated by a DM can be described by,
\begin{equation}
    E= A(1+g)e^{i\phi}.
    \label{eq:pupil_ef}
\end{equation}
Here $A$ is the pupil function, $\phi$ the DM surface shape and $g$ the aberration function, which includes both amplitude and phase aberrations. Propagation and apodization with any real or complex mask are linear operators in electric field. This means that a complete optical system can always be described by a single linear transfer function $C$ that relates the incoming electric field (pupil plane), $E$, with the final focal plane electric field. The focal-plane electric field, $E_f$, that corresponds to the pupil electric field of Eq. \ref{eq:pupil_ef} is,
\begin{equation}
    E_f = C\left\{A(1+g)e^{i\phi}\right\}.
    \label{eq:fpef}
\end{equation}
The modulated electric field can be expanded by means of a Taylor expansion as,
\begin{equation}
    E_f =\prop{ A(1+g) \sum^{\infty}_{n=1} a_n \phi^n } = \sum^{\infty}_{n=1} a_n \prop{A(1+g)\phi^n} = \sum^{\infty}_{n=1} a_n E_n.
\end{equation} %$a_n$ are the coefficients of the Taylor expansion is done are $a_n$ that corresponds to the nth order expansion has been substituted by $E_n$.
Here, $a_n$ are the Taylor expansion coefficients of terms $E_n$. The focal plane intensity then follows as,
\begin{equation}
    I = \left|\sum^{\infty}_{n=0} a_n E_n \right|^2 = \sum^{\infty}_{n=0}\left|a_n E_n \right|^2 + 2\sum^{\infty}_{n\neq m} \Re\left\{ a_n a_m^{\dagger} E_n E_m^{\dagger}  \right\}
    \label{eq:fpi}
\end{equation}

%This can now be used to define the focal plane electric fields, $E_S = C\eab$, $E_P=C\eab (e^{i\phi}-1)$. The intensity is,
%\begin{equation}
%    I_n= I_S + I_P + 2\Re\left\{E_P E_S^\dagger \right\}
%    \label{eq:fpi}
%\end{equation}

% In most cases both the aberrations and the probe are small, $g \ll 1$ and $\phi_n \ll 1$. This means that the last term can be neglected. The probed electric field, after a first order expansion of the probe, in the pupil is,
%\begin{equation}
%    E_n = A(1 + g + i\phi_n).
%    \label{eq:pupil_ef}
%\end{equation}

\subsection{Reconstructing the focal plane electric field}
The focal plane electric field has to be measured before EFC can be applied. Usually, the electric field is measured by adding a phase diversity on the DM (e.g., pair-wise probing) or by modifying the system in such a way that the electric field can be determined directly (e.g., the Self-Coherent Camera (SCC) \citep{baudoz2006scc}). Both approaches use post-processing to remove incoherent components from the intensity image and reconstruct the modulated part of the focal plane electric field.

With phase diversity this can be achieved by probing the focal plane with two probes with opposite sign, which leads to two measurements. A sign change of the phase probe switches the sign of all the uneven terms of the expansion in Eq. \ref{eq:fpef}, while the even order terms stay the same. For the focal plane intensity the effect is more complicated because the electric field components are mixed. The incoherent term, $|E_n|^2$, is always even and therefore will not be influenced by a sign change. The sign change for the cross terms depends on whether sum of $n$ and $m$ is odd, because $a_n a_m \rightarrow (-1)^n a_n (-1)^m a_m$ under a sign change. When $n$ and $m$ are odd the product of $(-1)^{n+m}=-1$, and when the sum of $n$ and $m$ is even $(-1)^{n+m}=1$:
\begin{equation}
\begin{split}
    I^{\pm} &= \sum^{\infty}_{n=0}\left|a_n E_n \right|^2 + 2\sum^{\infty}_{n\neq m, n+m=\mathrm{even}} \Re\left\{ a_n a_m^{\dagger} E_n E_m^{\dagger}  \right\} \\ & \pm 2\sum^{\infty}_{n+m=\mathrm{odd}} \Re\left\{ a_n a_m^{\dagger} E_n E_m^{\dagger}  \right\}
\end{split}
\end{equation}
The difference in intensity between the two probes is,
\begin{equation}
    I^+ - I^- = 4\sum^{\infty}_{n+m=\mathrm{odd}} \Re\left\{ a_n a_m^{\dagger} E_n E_m^{\dagger}  \right\}.
\end{equation}
This is the full expansion including all high-order terms. A common assumption is that the DM probe (and correction) are small and well within the linear regime ($\phi \ll 1$). In this limit only the lowest order terms have to be retained. With the added constraint that $n+m=\mathrm{odd}$, the first order that does not vanish is $n=0$ and $m=1$. From this the intensity difference follows as,
\begin{equation}
\begin{split}
    I^+ - I^- &= 4 \Re\left\{a_0 a_1^{\dagger} E_0 E_1^{\dagger} \right\}  + \bigo{\phi^3} \\
    &\approx 4 \Re \left\{ i \prop{A(1+g)} \prop{A(1+g)\phi}^{\dagger} \right\}.
\end{split}
\end{equation}
The image difference is a third-order approximation because the next terms are $(n=1, m=2)$ and $(n=0, m=3)$, both scale as $\phi^3$. The zeroth-order term can be recognized as the focal plane speckles without any modulation, $E_S = \prop{A(1+g)}$. And the second part of the difference image is the focal plane probe electric field, $E_P = i\prop{A(1+g)\phi}^\dagger$. The main difference between this expansion and those in the past is that the $g$ term has not been neglected for the probe electric field. The image difference can be rewritten into a more elegant form using vector products,
\begin{equation}
I^+ - I^- = 4
\begin{bmatrix} \Re\{E_P\} & \Im\{E_P\} \end{bmatrix}
\begin{bmatrix} \Re\{E_S\} \\ \Im\{E_S\} \end{bmatrix}.
\label{eq:single_probe}
\end{equation}

%\begin{equation}
%    \Delta E_P = 2i C\eab \sum^{\infty}_{n=1} \frac{i^{2n}
%    }{(2n+1)!} \phi^{2n+1}.
%\end{equation}
%The difference in electric field only depend on the uneven terms, because all even terms have been cancelled. The intensity, $I_P$, of both probes is also the same because the only difference is the change is sign of the linear term. The cross-talk term also changes when the sign of the probe changes. An inspection of the cross-talk electric field definition shows that it is very similar to the probe electric field,
%\begin{equation}
%    \Delta E_C = 2C\left\{Agi\phi\right\}.
%\end{equation}
%The crosstalk intensity, $I_C$, also does not change for a plus and minus probe.

%The positive probe intensity,
%\begin{equation}
%    I_n^+= I_S + I_P + 2\Re\left\{E_S E_P^\dagger \right\},
%\end{equation}
%and the negative probe intensity,
%\begin{equation}
%    I_n^-= I_S + I_P - 2\Re\left\{E_S E_P^\dagger \right\}.
%\end{equation}
%The difference between the two images is
%\begin{equation}
%    I_n^+ - I_n^-= 4\Re\left\{E_P^\dagger E_S \right\}.
%\end{equation}
At least two difference measurements with diverse enough probes are necessary to fully reconstruct the electric field because the electric field contains both a real and imaginary part. If we use $N$ pair-wise probes and separate the problem in the real and imaginary parts Eq. \ref{eq:single_probe} can be rewritten as a linear matrix problem,

\begin{equation}
\label{eq:full_system}
\begin{bmatrix}
I_1^+ - I_1^-\\ 
\vdots\\ 
I_N^+ - I_N^-
\end{bmatrix} = 4
\begin{bmatrix}
\Re\{E_1\} & \Im\{E_1\} \\ \vdots & \vdots \\ \Re\{E_N\} & \Im\{E_N\} \end{bmatrix}
\begin{bmatrix} \Re\{E_S\} \\ \Im\{E_S\} \end{bmatrix}.
\end{equation}
With $E_i$ the electric field of probe $i$. This can be simplified by defining a system matrix,
\begin{equation}
M= 4 \begin{bmatrix}
\Re\{E_1\} & \Im\{E_1\} \\ \vdots & \vdots \\ \Re\{E_N\} & \Im\{E_N\} \end{bmatrix},
\end{equation}
a measurement vector,
\begin{equation}
\delta = \begin{bmatrix} I_1^+ - I_1^- \\ \vdots\\ I_N^+ - I_N^- \end{bmatrix},
\end{equation}
and a state vector,
\begin{equation}
\psi_S = \begin{bmatrix} \Re\{E_S\} \\ \Im\{E_S\} \end{bmatrix}.
\end{equation}
Substituting these definitions into Eq. \ref{eq:full_system} results in,
\begin{equation}
    \delta = M \psi_S
    \label{eq:system_matrix}
\end{equation}
The electric field is estimated by solving the regularization least-squares problem,
\begin{equation}
    \psi_S = \argmin_{\psi_S} |\delta - M \phi_S|^2 + \mu |\phi_S|^2.
\end{equation}
\edited{The regularization parameter $\mu$ is used to penalize solutions with large phases. The solution of this optimization problem is}
\begin{equation}
    \psi_S = \left(M^TM + \mu I\right)^{-1}M^T \delta.
    \label{eq:pwp_recon}
\end{equation}
%For the SCC the derivation of the reconstruction of the focal plane electric field can be found in \citep{mazoyer2013labverification}.

\subsection{The Electric Field Conjugation controller}
In the framework of explicit EFC, the DM correction is assumed to create a focal plane electric field similar to that of the probes. The phase of the DM is described as a linear combination of basis modes, $\phi_{DM}=B\alpha$ with $B$ the basis transformation and $\alpha$ the basis coefficients. The electric field that the DM creates is,
\begin{equation}
    E_{DM} = i\prop{\phi_{DM}} = i\prop{B \alpha}.
\end{equation}
The focal plane electric field that the DM creates is similar to the electric field that the probes create (Eq. \ref{eq:single_probe}), however the $g$ term is usually neglected because an aberration free system is assumed. The basis transformation and the propagation matrix can be combined into a single transfer function $G$, which describes the electric field created by each DM mode. To determine the control command EFC again sets up a regularized least-squares problem,
\begin{equation}
    \alpha = \argmin_{\alpha} |\psi_S + G \alpha|^2 + \lambda |\alpha|^2.
\end{equation}
\edited{The regularization parameter $\lambda$ is used to penalize solutions with large actuator responses. This effectively penalizes solutions with a strong negative impact on Strehl.} The reconstructed control commands are then,
\begin{equation}
    \alpha = -\left(G^TG + \lambda I\right)^{-1}G^T \psi_S.
\end{equation}
The reconstruction (Eq. \ref{eq:pwp_recon}) and control can then be combined into a single equation,
\begin{equation}
    \alpha = -\left(G^TG + \lambda I\right)^{-1}G^T \left(M^TM + \mu I\right)^{-1}M^T \delta.
    \label{eq:EEFC}
\end{equation}
This equations makes it clear why this method is called explicit EFC. The focal plane electric field is first explicitly reconstructed before the optimal control command is calculated.

\subsection{The iEFC controller}
Here we derive the iEFC controller, which has the benefit that no model knowledge of your optical system is required. The basis of iEFC starts with Eq. \ref{eq:system_matrix} which shows how an electric transforms into measured difference images. iEFC uses this equation to transform the DM commands into modulated difference images,
\begin{equation}
    \delta = MG\alpha = Z\alpha.
\end{equation}
This shows that there is a single response matrix, $Z=MG$, that relates the effect of the DM modal coefficients with the modulated difference images. In EFC, the focal plane speckles are removed by injecting the opposite electric field, which results in destructive interference and therefore a dark hole. This is not possible with iEFC because the electric field is not reconstructed, only the difference images are accesible. This issue can be sidestepped by minimizing the difference images themselves, which also results in minimizing the electric field because the difference images are a linear proxy of the focal plane electric field (see Eq. \ref{eq:system_matrix}). The control optimization problem of iEFC is, 
\begin{equation}
    \alpha = \argmin_{\alpha} |\delta + Z \alpha|^2 + \lambda |\alpha|^2,
\end{equation}
which has as solution,
\begin{equation}
    \alpha =  -(Z^TZ +\lambda I)^{-1}Z^T \delta.
    \label{eq:IEFC}
\end{equation}
This controller is fundamentally different from the EFC controller (Eq. \ref{eq:EEFC}). The electric field is not estimated as an intermediate product, however it is still minimized. This is why this controller is called the implicit EFC controller. An advantage of iEFC is that it is not necessary to optimize the electric field regularization parameter $\mu$. The response of the DM modes in the modulated intensity already regularizes the problem because the response of a DM mode usually extends over several focal plane pixels. The iEFC controller was derived in the context of pair-wise probing, however any measurement that is sensitive to the modulated electric field can be used for $\delta$ (such as the side-band signal of the SCC).

iEFC is easily extended to multiwavelength measurement by concatenating the difference images for each wavelength,
\begin{equation}
    \delta_{\mathrm{poly}} = \begin{bmatrix} \delta_{\lambda_1} \\ \vdots \\ \delta_{\lambda_L}\end{bmatrix}.
\end{equation}
The controller can also be extended to include a focal plane weight map or a different regularization. Most EFC algorithms use Tikhonov regularization or a modal truncation \citep{ruane2022broadband}. Tikhonov regularization adds a penalty by including the quadratic norm of the control vector. However, other regularizations can also be used such as a smoothness regularizer \citep{engl1996regularization}. Adding all three extensions results in the following optimization problem,
\begin{equation}
    \alpha = \argmin_{\alpha} \left(\delta_{\mathrm{poly}}^T + \alpha^T Z_{\mathrm{poly}}^T \right) Q \left(\delta_{\mathrm{poly}} + Z_{\mathrm{poly}} \alpha\right) + \alpha^T R\alpha.
\end{equation}
Here $Q$ is the focal plane weight matrix, $R$ is the regularization matrix and $\delta_{\mathrm{poly}}$ is the multi-wavelength measurement vector. This cost function can be recognized as the cost function for a Linear Quadratic Regulator (LQR). The controller that minimizes this cost function is,
\begin{equation}
    \alpha = -\left(Z_{\mathrm{poly}}^T Q Z_{\mathrm{poly}} + R\right)^{-1} Z_{\mathrm{poly}}^T Q \delta_{\mathrm{poly}}.
\end{equation}

\subsection{Broadband iEFC}
High-contrast imaging is pushing toward broader spectral bandwidths for characterization of exoplanet atmospheres. Therefore, it is important to be able to work in broadband light. iEFC assumes a linear relation between the DM and modulated intensity. This relation is not removed when broadband light is used because the broadband signal will incoherently add, which means that the DM response will be averaged over the input spectrum. The spectrum averaged response of the DM is, 
\begin{equation}
    \delta_{bb} = \int_{\lambda_0}^{\lambda_1} F\left( \lambda \right) Z\left(\lambda\right) \alpha \mathrm{d}\lambda.
\end{equation}
The difference images are now the broadband integrated difference images, $\delta_{bb}$. And the input spectrum is $F\left(\lambda\right)$. For broadband light, the DM response matrix is chromatic, but the modal coefficients are not chromatic. Effectively, the DM response is averaged over the spectrum and this can be made explicit by using the total flux within the passband
\begin{equation}
    \delta_{bb} = F_T \langle Z\rangle_F \alpha.
\end{equation}
Here $F_T$ is the total flux. The broadband response matrix can be calibrated in exactly the same way as the monochromatic version. The control matrix can be found by substituting the broadband response matrix and the broadband difference images into the iEFC solution (\ref{eq:IEFC}).

\subsection{Calibration of the iEFC response matrix}
\subsubsection{Calibrating a low jitter system}
For iEFC the response of the DM is directly related to a measurable, which means that the response matrix can be directly calibrated and it is not necessary to model it. The response matrix can be measured by poking the DM modes. Before wavefront control the response of iEFC still includes the aberrated electric field. The measured differences when adding mode $j$ with amplitude $a$ to the DM is,
\begin{equation}
    \delta_j = M\psi_S + Z_j a.
\end{equation}
Here $Z_j$ is the column $j$ of matrix $Z$. To remove the static electric field, both a positive and negative poke needs to be applied. So for every mode a positive poked measurement is taken,
\begin{equation}
    \delta_j^+ = M\psi_S + Z_j a,
\end{equation}
and a negative poked measurement,
\begin{equation}
    \delta_j^- = M\psi_S - Z_j a.
\end{equation}
The difference between the $\delta$'s then results in twice the mode response,
\begin{equation}
    \Delta\delta_j = \delta_j^+ - \delta_j^- = 2Z_j a.
\end{equation}
This measurement is called the double difference image. The double difference is taken to remove all static aberrations from the response matrix. The final response matrix is constructed by repeating the double difference measurement for all $K$ DM modes that need to be controlled.
\begin{equation}
\label{eq:pushpullcal}
    \hat{Z} = \frac{1}{2a}\begin{bmatrix} \delta_1^+ - \delta_1^- \hdots  \delta_K^+ - \delta_K^- \end{bmatrix}.
\end{equation}
The control matrix is then found by substituting this into Eq. \ref{eq:IEFC}.

%The response of the pair-wise probing under broadband light is an incoherent sum of the monochromatic response. This is fundamentally a different problem from the multi-wavelength measurements. 

%The measured intensity difference, under the first-order approximation, is
%\begin{equation}
%    \delta = \int_{\lambda_0}^{\lambda_1} F\left(\lambda \right) M\left( \lambda \right) \psi_S \left(\lambda\right)  \mathrm{d}\lambda.
%\end{equation}
%Both the system matrix and the electric field state vector are wavelength dependent and multiplied by the spectrum $F\left(\lambda\right)$.

\subsubsection{Calibrating a system with jitter}
The calibration method described above has reduced Signal-to-Noise if the system has a significant amount of jitter. The $Z$ matrix registers the DM and the pixel responses. If the PSF has some tip/tilt offset, the wrong DM to pixel map is calibrated. This effect can be reduced by integrating over typical coherence times of the jitter. A downside is that this approach will increase the duration of the calibration. A solution is to apply many random linear combinations of modes on the DM. The response matrix can then be reconstructed if enough random combinations are measured. The best calibration is made when the response of a positive and negative version of the random combination are taken. The double difference images are necessary to remove the intrinsic speckles that are present in the system. The response matrix can then be created with

\begin{equation}
\label{eq:randomcal}
    \hat{Z} = \Delta\delta V^T \left(V V^T + \gamma I \right)^{-1}.
\end{equation}
The response matrix $Z$, depends on both the random inputs $V$ and the recorded double difference images $\Delta\delta$. The identity matrix $I$ is used as regularization matrix with $\gamma$ as the strength of the regularization. 

\subsection{Summary of the iEFC algorithm}
The iEFC algorithm can be summarized as follows. \editedLC{First, the focal plane region that will be nulled is selected. Then the Fourier modes corresponding to that region with an additional 1 $\lambda / D$ border are created. Oversizing the area that the Fourier modes span reduces edge effects at the border of the nulled region. The sensing probes are created as single actuator probes. The most optimal probes have a separation that is in the direction of the dark hole. The chosen Fourier modes and probing modes can then be used to measure the interaction matrix using either equation \ref{eq:pushpullcal} or equation \ref{eq:randomcal}. After calibrations, the iEFC algorithm takes a set of pair-wise probing measurements and then calculates and applies a correction. This is repeated until the dark hole has converged or a satisfactory contrast has been reached. If the contrast stagnates the response matrix can be recalibrated around the new state.}
%\begin{itemize}
%  \item Identify the region that the user wants to null.
%  \item Create all the Fourier modes that correspond \edited{to} the region with an additional 1 $\lambda / D$ border. Oversizing the area that the Fourier modes span will help with reducing edge effects at the border of the nulled region.
%  \item Select the single actuator probes. The most optimal probes have a separation that is in the direction of the dark hole.
%  \item Measure the interaction matrix using either equation \ref{eq:pushpullcal} or equation \ref{eq:randomcal}.
%  \item Take pair-wise probing measurements.
%  \item Apply a correction step.
%  \item Repeat the last two steps until the dark hole has converged or a satisfactory contrast has been reached.
%  \item \edited{If the contrast stagnates the response matrix can be recalibrated around the new state.}
%\end{itemize}

%\begin{algorithm}
%\caption{Digging a dark-hole with iEFC}\label{alg:cap}
%\begin{algorithmic}

%\State $i \gets 10$
%\If{$i\geq 5$} 
%    \State $i \gets i-1$
%\Else
%    \If{$i\leq 3$}
%        \State $i \gets i+2$
%    \EndIf
%\EndIf 
%\end{algorithmic}
%\end{algorithm}

\begin{figure}[ht]
\includegraphics[width=\columnwidth]{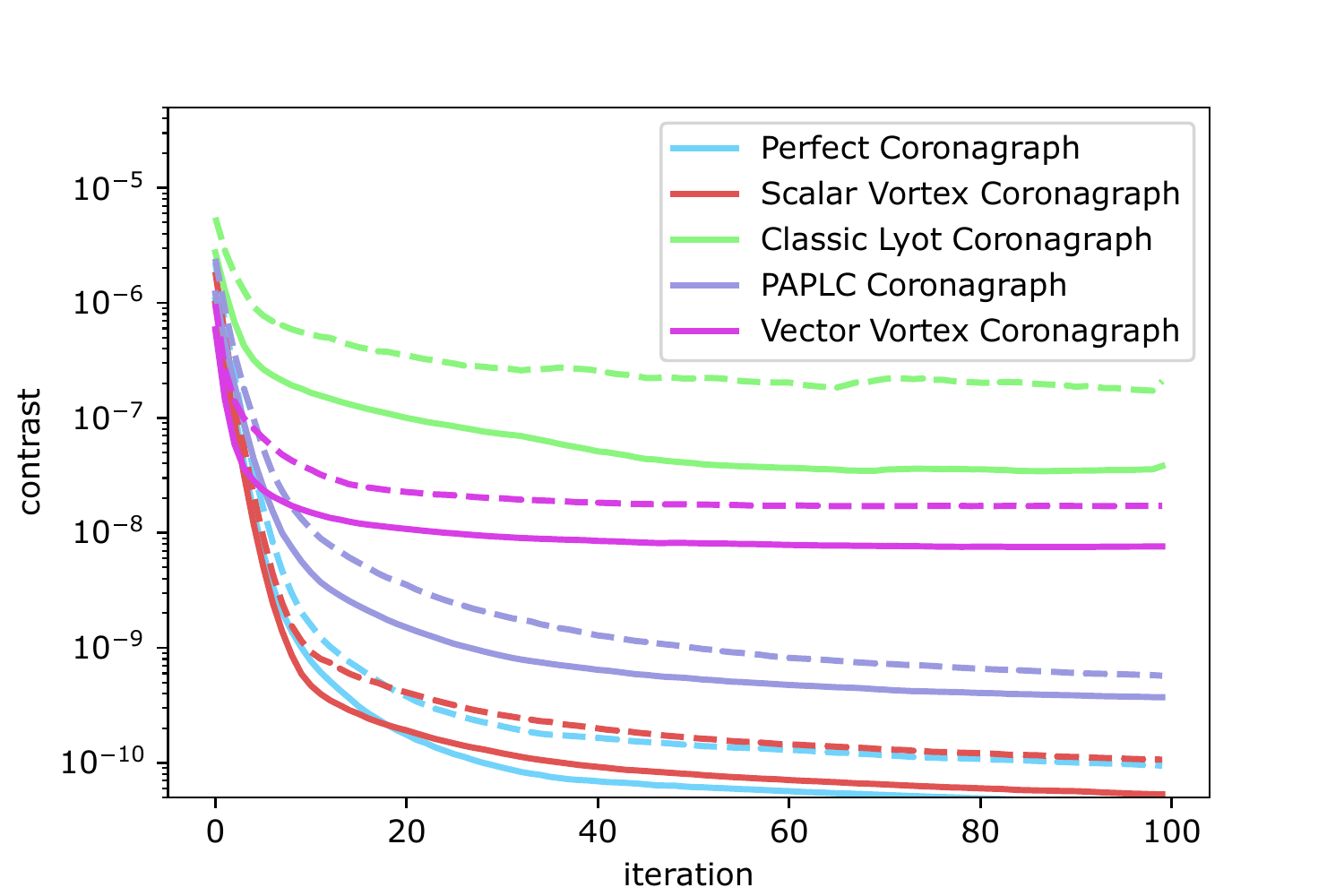}
\caption{Contrast as a function of iteration. Each color represents a different coronagraph. The solid lines shows the median contrast within the dark hole. \edited{The dashed lines correspond to the 84th quantile, which is the 1-$\sigma$ upper confidence bound. The Perfect, Vortex and PAPLC coronagraphs rapidly reach a contrast of $10^{-9}$ and then slowly converge to $10^{-10}$. The Classic Lyot and the Vector Vortex Corongraph converge to $\sim10^{-8}$.} The iEFC control\edited{ler} is also stable because even after 100 iterations the contrast \edited{does not diverge}.}
\label{fig:closed_loop_timeseries}
\end{figure}

\section{Simulations}

The iEFC algorithm is explored numerically in this section. The python module High Contrast Imaging for Python (HCIPy) \citep{por2018hcipy} is used for all simulations in this section. The simulations in this section uses a somewhat idealized system. A clear aperture with a single 50x50 DM is simulated. The coronagraphs that require a Lyot stop use a clear aperture with a 5\% under sizing. Single neighboring actuator probes are used for the pair-wise probing \citep{potier2020comparing}. These probes are convenient probes because they span the full control region. The axis along which the two actuators lie has to be chosen in the same direction as the orientation of the dark hole \citep{potier2020comparing}. The actuators of choice in these simulations are $(12, 25)$ and $(11, 25)$. Only a single DM is available in the considered system because most ground-based xAO systems only have a single DM, which limits the speckle control to one-sided dark holes. The considered dark hole runs from $2\lambda/D$ to $18 \lambda/D$ in the horizontal direction, and from $-15 \lambda/D$ to $15\lambda/D$ in the vertical direction. In theory, a larger dark hole can be created because the system has 50 actuators across the pupil. The focal plane image is sampled with 3 pixels per $\lambda/D$ at the central wavelength and has a field of view of $50 \lambda/D$ in diameter.

\begin{figure*}[ht]
\centering
\includegraphics[width=\textwidth]{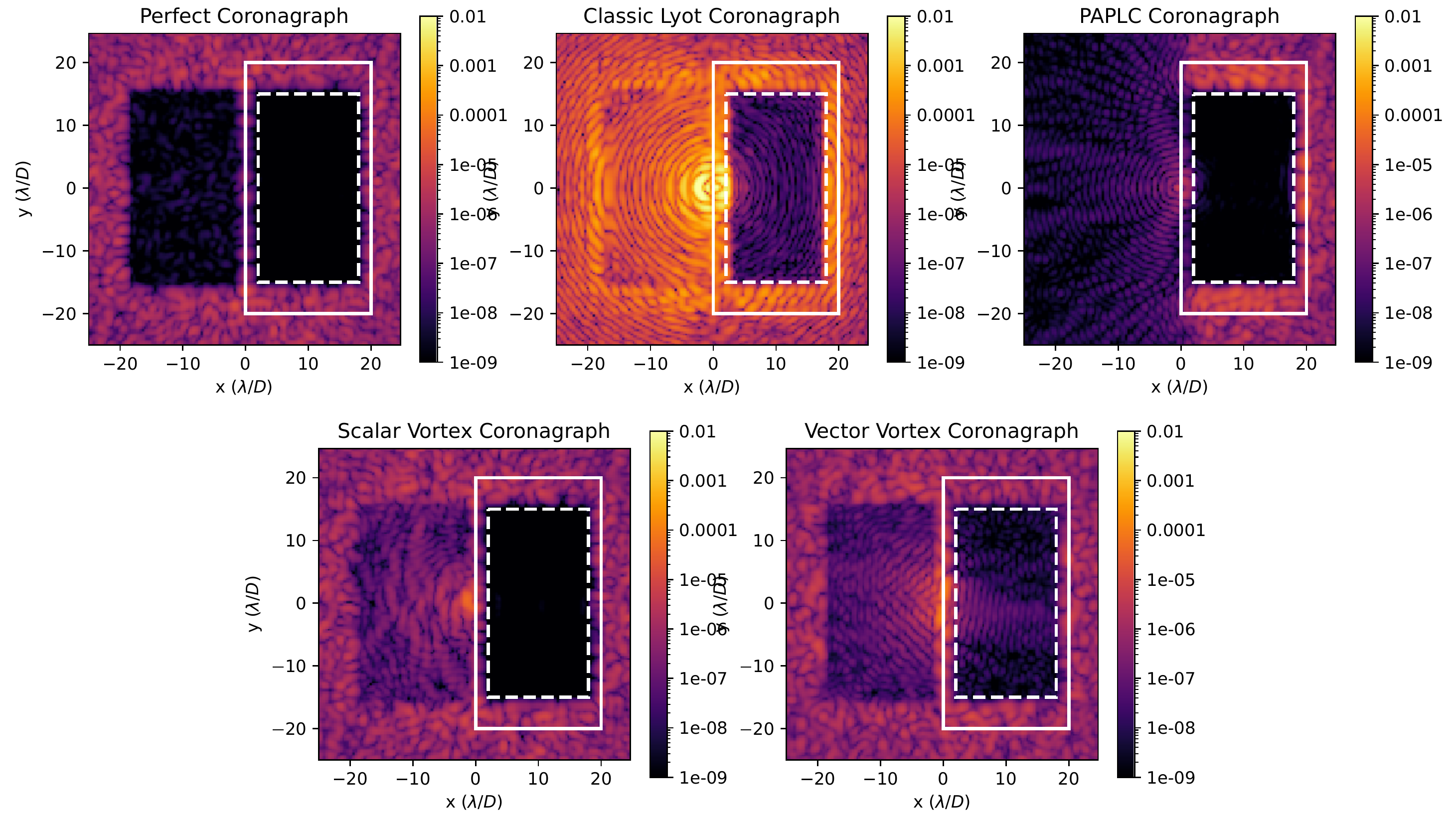}
\caption{Post-coronagraphic focal plane for various coronagraphs. The solid rectangles show the area that is used to sense the aberrations while the dashed area shows the area that is controlled. For the scalar corongraphs a median contrast below $10^{-9}$ is reached. Only the vector vortex coronagraph is not able to reach such a deep contrast for an unpolarized input beam.}
\label{fig:coronagraph_focal_plane}
\end{figure*}

Several aspects of the iEFC algorithm are explored. The robustness of the algorithm is demonstrated by controlling the focal plane speckles with various coronagraphs. We also show that a dark hole can be created in large spectral bandwidths and the effect of measurement noise both in the actual closed-loop measurements and the calibrations are analyzed. Finally, iEFC is also demonstrated behind an xAO system with atmospheric turbulence.

\subsection{Coronagraph comparison}
The first set of simulations explore the implementation of iEFC with different coronagraphs. The considered coronagraphs are the perfect coronagraph (PC) \citep{cavarroc2006fundamental, guyon2006coronagraph}, the \edited{Scalar and Vector Vortex Coronagraph (SVC/VVC)} \citep{mawet2005annular,foo2005optical}, a Classic Lyot Coronagrah (CLC) \edited{with a circular mask of 2 $\lambda/D$ in radius} and the recently developed knife-edge Phase Apodized Lyot Coronagraph (PAPLC) \citep{por2020phase}. The system matrix is calibrated for each coronagraph with instrumental speckles included. This will show that it is not necessary to calibrate around a perfect dark hole. The speckles are created by an out-of-pupil phase screen that has a -2 power-law distribution and a total power of $\lambda/10$ peak-to-valley. These amounts of NCPA are quite representative of modern xAO systems, SPHERE has 50 nm \citep{beuzit2019sphere} of residuals phase aberrations while MagAO-X has less than 30 nm \citep{van2021characterizing}. The phase screen is placed out of the pupil plane to induce some amount of amplitude aberrations.

The contrast as a function of iteration for all coronagraphs is shown in Figure \ref{fig:closed_loop_timeseries}. The final dark hole corresponding to each coronagraph can be seen in Figure \ref{fig:coronagraph_focal_plane}. The results show that the precise details of the instrument and the coronagraph do not matter for iEFC. The focal plane dark-hole is minimized regardless of the implementation. The median contrast of the PAPLC dips below $10^{-9}$ while the \edited{SVC} and PC go below $10^{-10}$. The contrast of the VVC plateaus at $\approx10^{-8}$. This is due to the differential effect that the VVC has on the two circular polarizations. The VVC introduces the vortex phase pattern with geometric phase which imparts opposite phase on the two circular polarization states. The DM can only cancel one of the two states, leaving residual speckles from the other polarization. This is an known phenomenon \citep{mendillo2021dual}, that is usually alleviated by adding polarizers in the system that remove one of the two states. This effectively reduces the VVC to a scalar VC. However, this also showcases the model free approach of the iEFC algorithm because no knowledge of polarization was added, and a dark hole could still be created. The contrast of the VVC is not sufficient for space-based mission, but would be deep enough for ground-based systems.

The contrast of the CLC is limited by the intrinsic design of the LC. The CLC does not completely remove the diffraction pattern for an aberration-free system. This means that the iEFC algorithm will need to remove static diffraction patterns. This is quite difficult compared to the PC and VC where wavefront control only needs to correct for aberrations. This limits the CLC dark hole performance to roughly $\sim 10^{-7}$.

\subsection{Broadband control}
\begin{figure*}[ht]
\includegraphics[width=\textwidth]{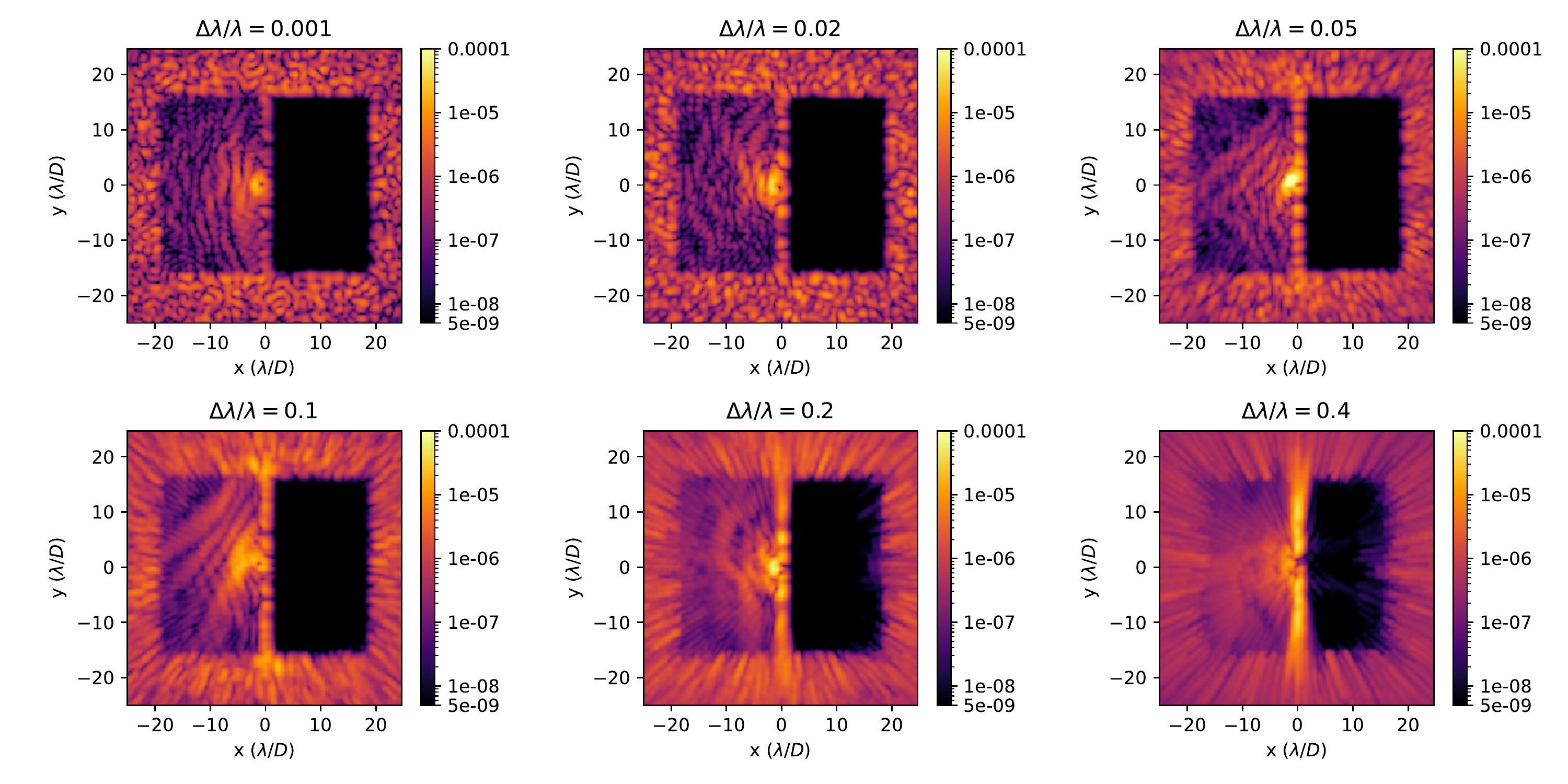}
\caption{Coronagraphic focal planes after iEFC for several different bandwidths. The lowest bandwidth is effectively monochromatic and the largest bandwidth is 40\%. A rectangular dark hole was created that spans the area from -15 $\lambda/D$ to 15 $\lambda/D$ in the vertical direction and from 2 $\lambda/D$ to 20$\lambda/D$ in the horizontal direction. The majority of the controlled area stays below $5\cdot10^{-9}$ at all bandwidths. At large spectral bandwidths the performance at the edges start to degrade.}
\label{fig:bandwidth_performance}
\end{figure*}

There is a lot of interest in characterize the atmospheres of exoplanets with direct imaging. The accuracy of the atmospheric retrieval depends on the spectral bandwidth that is used \citep{damiano2022reflected}, which drives direct imaging systems to work with very broad spectral bandwidths. The coronagraphic focal plane after wavefront control for a range of spectral bandwidths is shown in Figure \ref{fig:bandwidth_performance}. All simulations here use a Vortex Coronagraph that because this is in theory an achromatic coronagraph. \edited{The VC that we consider here uses an achromatic phase mask. This can either be created by using the VVC in a single polarization (with a 50\% loss of light) or by stacking multiple chromatic SVC masks. The precise implementation of the VC does not matter for test because its the chromatic speckle sensing and control that is tested.} Each system matrix is calibrated before the wavefront control. The results show that iEFC can handle spectral bandwidths up to 40 \%. This is comparable to the bandwidth of the J and H band together (39.1 \%). The median contrast does degrade from below $10^{-9}$ in semi-monochromatic light to $\approx 10^{-8}$ in broadband light. This is not because the speckles can not be nulled, but because the size of the dark hole shrinks and grows with wavelength. The larger wavelengths have some residuals at 1 to 2 $\lambda/D$ which are further out at shorter wavelengths. These spatial frequencies are not controlled at the longer wavelengths, leading to reduced contrast. The opposite happens at the edge of the control region where the speckles of the shorter wavelengths are not controlled.

These simulations do not show how an actual instrument would respond because the precise chromatic behavior of speckles matters. An important component of the chromatic behavior is Fresnel diffraction, which changes the phase and amplitude of plane waves as they propagate. The precise strength and chromatic behavior depends on the distance the waves have to propagate to the pupil. Which means that the exact position of the DM and all other optics matter to get the correct chromatic behavior. Such an extensive modeling of instrumental effects is outside the scope of this work. In the simulations we use only a single DM, which means that it is not possible to correct complicated chromatic behavior. A solution to still create deep dark holes, even in the presence of strong chromatic behavior, is to use multiple DMs \citep{pueyo2009optimal, baudoz2018status}.

\subsection{Robustness against noise}
\begin{figure}[ht]
\includegraphics[width=\columnwidth]{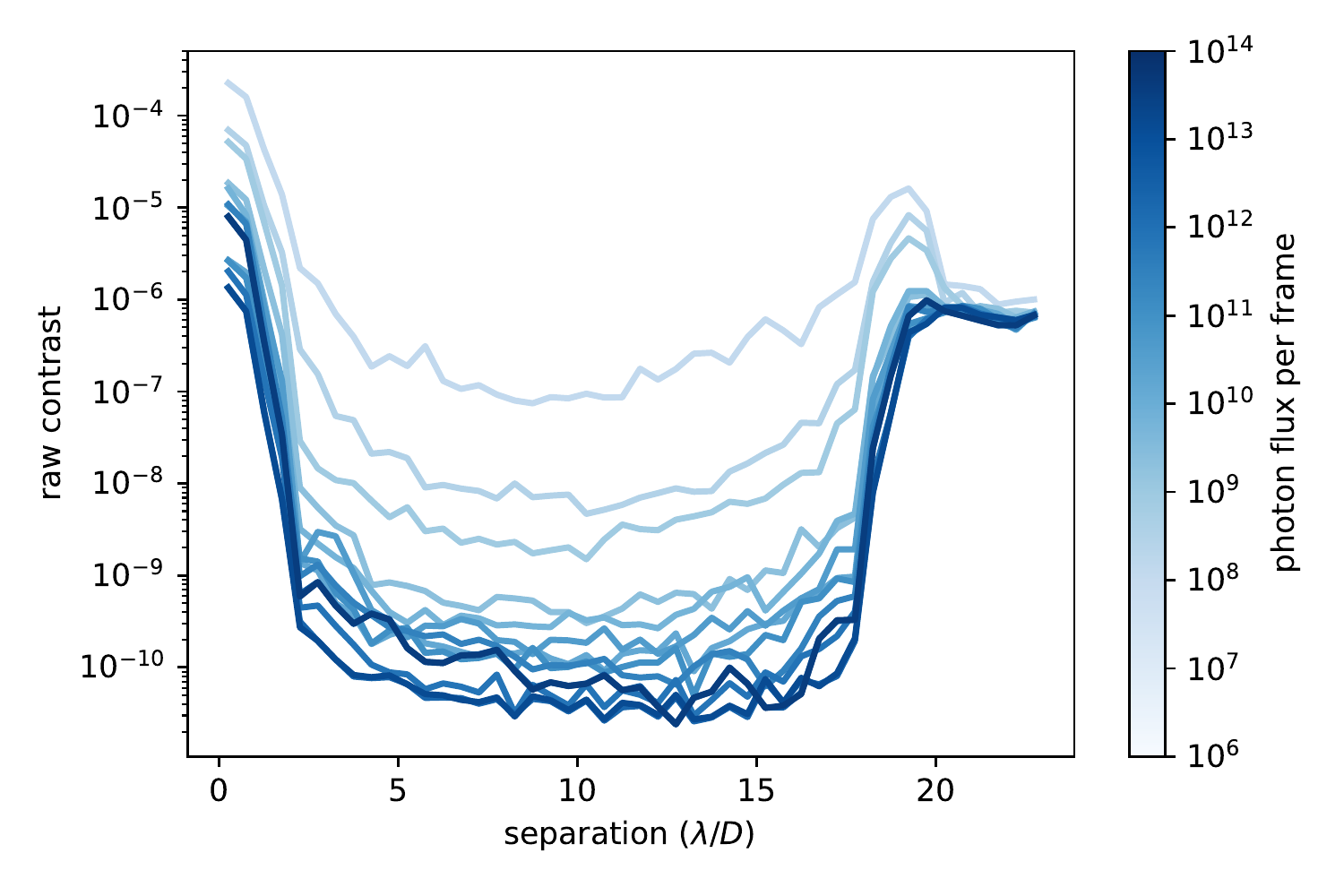}
\caption{Final achieved contrast curves as a function of calibration photon flux. The dark hole is a half-dark hole with an inner working angle of 1.5 $\lambda/D$. The dark hole with the lowest photon flux reaches a contrast of $10^{-7}$, while the dark hole with the highest photon flux reaches $3\cdot10^{-11}$.}
\label{fig:calibration_requirement}
\end{figure}

\begin{figure}[ht]
\includegraphics[width=\columnwidth]{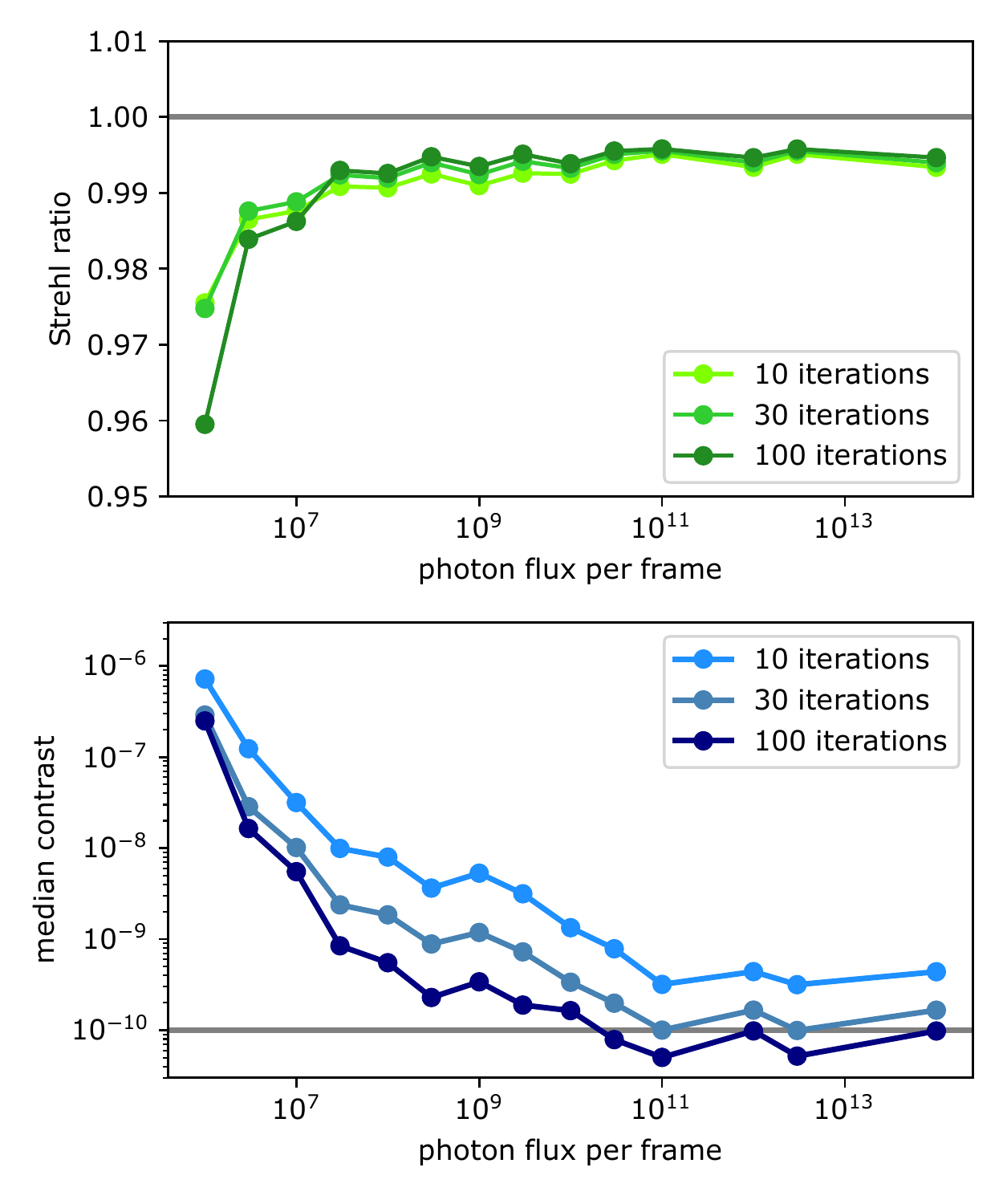}
\caption{\editedLC{Performance of iEFC as a function of photon flux.} The top figure shows the Strehl ratio as a function of calibration flux after closed-loop control and, the bottom figure shows the median contrast as a function of calibration flux. Both the Strehl ratio and contrast are measured after a number of closed-loop iterations. Each line corresponds to a different number of iterations. The brightest green (blue) shows the Strehl ratio (contrast) after 10 iterations while the darkest green (blue) shows the Strehl ratio (contrast) after 100 iterations. A large improvement is visible if enough photons are used to calibrate the system matrix. The achieved performance converges when roughly $10^{10}$ photons per calibration frame are used. }
\label{fig:snr_performance}
\end{figure}

The response matrix can be empirically calibrated in the framework of iEFC. When the response matrix is measured, measurement noise will affect the response matrix, putting a limit on the achievable contrast and correction speed. The impact of the measurement noise is investigated by creating the response matrix with varying photon flux levels \edited{for the Perfect Coronagraph}. A single mode requires eight images to calibrate the response with the double difference method. The photon flux per frame is defined as the photon flux for a single frame out these eight frames. The measured response matrix is then used in closed-loop without adding additional noise to the closed-loop images. This was done to disentangle calibration noise and measurement noise. The achievable contrast for varying photon flux is shown in Figure \ref{fig:calibration_requirement}. The photon flux is defined as the amount of photons entering the system before the coronagraph. A photon flux of at least $10^8$ photons per frame are required to achieve a contrast of $1\cdot 10^{-8}$. Higher contrast requires more photons but, this relationship does not follow the expected $\propto \sqrt{N}$ relationship. The median contrast in the dark hole and the Strehl ratio as a function of photon flux is shown in Figure \ref{fig:snr_performance}. The final contrast of $10^6$ photons/frame stagnates at $10^{-7}$ to $10^{-6}$, and a contrast of $10^{-9}$ is reached with $10^9$ photons. This indicates that there is a linear relation between final contrast and calibration flux. The final contrast converges around $10^{10}$ photons/frame. More photons than $10^{10}$ per frame does not lead to a deeper contrast, but it does lead to faster convergence. The final contrast is $10^{-10}$ after 100 closed-loop iterations with $10^{10}$ photons. With $10^{11}$ photons/frame only 30 iterations are necessary to achieve the same contrast. More photons than $10^{11}$ per frame did not lead to fast convergence or deeper contrast. The simulations also show that at least $10^8$ photons are required to achieve high Strehl ratio. At lower photon-flux levels the Strehl ratio will actually decrease in closed-loop operations. \edited{At low signal-to-noise ratio (S/N) the modes that do not make a big impact in contrast are also not measured well. This is an intrinsic property of iEFC because the response in the dark hole is measured for each calibrated mode. This means that at low S/N it is still possible to improve contrast by controlling the worst offenders. However, on average over all modes the total phase rms increases leading to lower Strehl. A smarter mode basis for the controller could solve this issue.}

\subsection{Atmospheric turbulence}
For ground-based high-contrast imaging there is an additional complication, caused by the post-AO atmospheric phase residuals. The post-AO atmospheric speckles change at very short timescales. Typical timescales are 1 to 10 ms. The varying AO speckles will not cancel in subsequent measurements in the pair-wise probing sequence if short exposures are used. This will lead to artificial NCPA signals \citep{singh2019active, potier2019exoplanet}. And at short exposures there might not be enough photons to measure the wavefront accurately enough. Both problems can be reduced by taking long exposures that average over multiple residual phase realizations (multiple speckle coherence times). The downside of long exposures is that they reduce the effective speed of the wavefront control loop. This trade-off between loop speed and photons noise is similar to the trade-off for normal AO.

%Another disadvantage of short exposures is that there may not be enough photons to measure the quasi-static speckles. 

%At short integration times the speckles change from measurement to measurement. This introduces noise in the pair-wise probing because the varying AO speckles do not cancel out, which lead to artificial NCPA signals \citep{singh2019active, potier2019exoplanet}. This effect can be reduced by taking exposures that average over multiple speckle coherence times.

%with long enough integration time soo that the have to be taken to average over enough phase realizations.
%Instantaneous (short) exposures cannot be used in combination with pair-wise probing due to the non-perfect cancellation of the atmospheric aberrations in .

%This will separate to generate a one minute time series.%The effects of different focal-plane wavefront sensing loop speeds on the residual quasi-static speckles.
This trade-off is explored by simulating a 60 seconds sequence of residual AO phase screens. The phase screens are generated by an AO system similar to MagAO-X \citep{males2022magao}. The system has 50 actuators across the pupil, with a total of 1600 controlled modes. The AO loop ran at 2 kHz with a gain-optimized integral controller. The simulation generated a total of 120000 frames. \edited{A Perfect Coronagraph is used again for the simulations of this system.}
%The residual AO speckles introduce a significant amount of noise for short exposures because the speckle pattern changes from frame to frame. While longer exposures have less influence from the AO speckles because they are averaged out during the exposure.

\edited{Longer} exposures average the AO speckles and should lead to better correction of the quasi-static speckles \edited{because multiple speckle realizations are averaged and therefore less noise will propagate into the NCPA reconstruction}. \edited{A loop running on short exposures will have a larger error because of the atmospheric speckle noise. However, there are many more iterations to control the speckles if the same amount of total exposure time is used.} For example, if the short exposures use 5 ms and the long exposures 1 s, then the short exposures have 200 times more iterations. So, even if the short exposures create more noise on a per measurement basis, it could be that after many control iterations the final contrast is better. Multiple frames are combined such that the total integration time is 1 second \edited{for the evaluation of the final contrast. This makes for a fairer comparison between different control rates.}
%In these simulations, we look at a total exposure time of 1 second.

\begin{figure}[ht]
\includegraphics[width=\columnwidth]{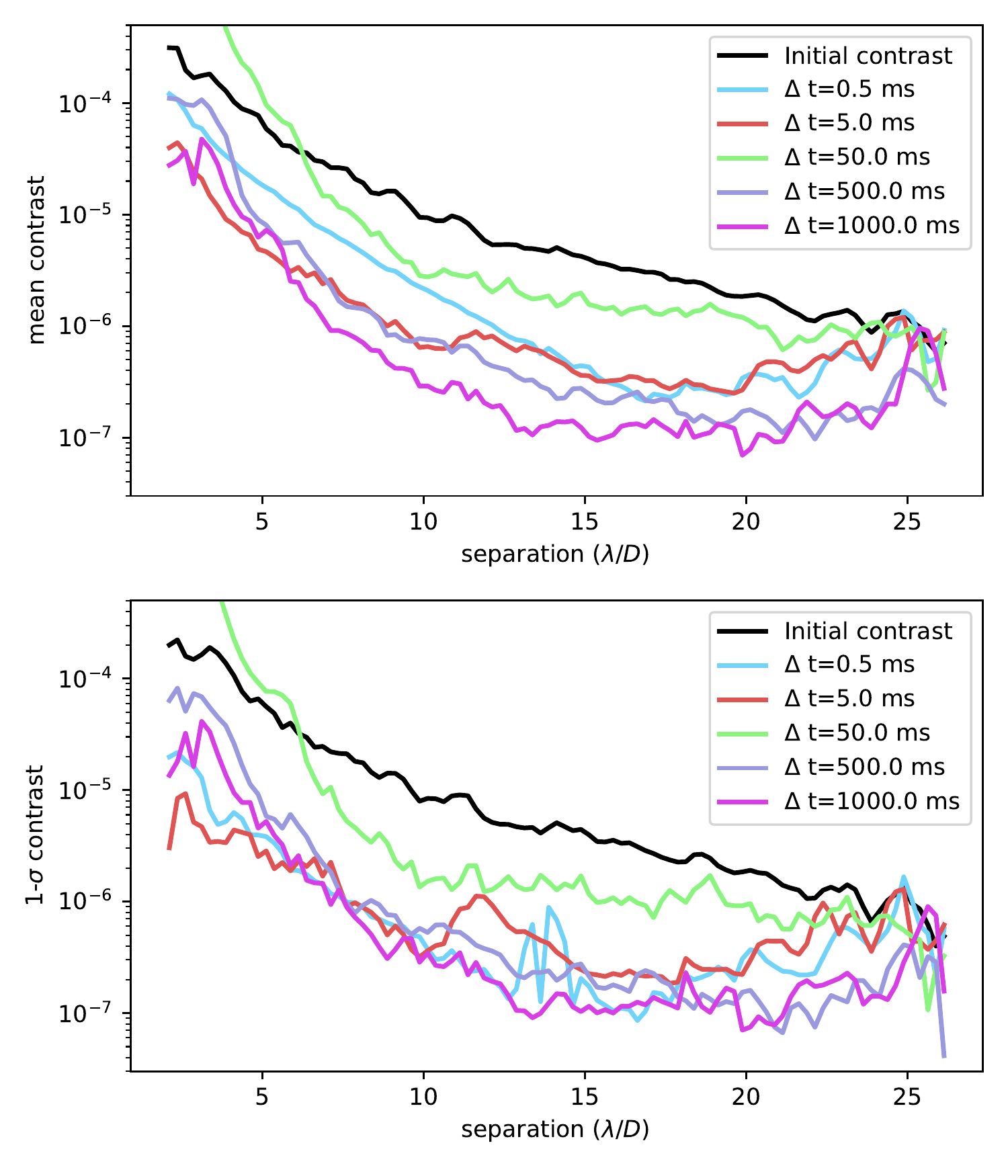}

\caption{Spatial radial profile of the quasi-static speckles in the dark hole after one second of closed-loop operation. The top figure shows the mean radial profile in the dark hole for different exposure times and, the bottom figures shows the radial standard deviation for each angular separation. The black line corresponds to the input profile for the quasi-static speckles. The shortest exposure lead to a smooth residual halo, which shows how the atmospheric halo leaks into the quasi-statics. The results for 50 ms exposures always diverge regardless of closed-loop gains and regularization.}

\label{fig:postao_sim}

\end{figure}
 %These curves do not include the atmosphere halo because we are only interested in the speckles from NCPA.
The strength of the quasi-static speckles after wavefront control are shown in Figure \ref{fig:postao_sim}. The atmospheric halo has been subtracted from the images to highlight the effects of the wavefront control on the quasi-static speckles. The black line shows the initial mean and standard deviation of the contrast curves. The behavior as a function of exposure time is quite complex. The shortest exposure time at 0.5 ms result in a smooth halo. This is because the residual AO speckle halo does not average out between exposures on these short time scales, which causes the halo to bleed into the quasi-static correction. The exposure time is relatively short compared to the evaluation time of 1 second. \edited{The evaluation exposure of 1 second} contains many realization which smooths the speckles and it creates a smooth NCPA halo. This is quite clear in the top figure of \ref{fig:postao_sim}. The 1-$\sigma$ contrast is still quite low because that residual halo is so smooth. The contrast curves become less smooth and deeper for longer exposure times. The longest exposure time creates the deepest dark hole. The 50 ms exposure time was never stable and diverged for every combination of feedback gain and regularization that we tried. This is most likely because this exposure time is closest to the coherence time of the AO residuals, which means that the temporal speckle noise from changing atmospheric residuals is the largest. The longest exposure times have the deepest contrast but, their $1-\sigma$ contrast is nearly equal to their mean profile. This is an effect from small number statistics because the longest exposure times also have the fewest realizations due to the finite time series. For the longest time series we only had 15 independent measurements. Due to the feedback gain of 0.4 only the last few frames are useful for estimating the performance. The first few frames still have all the speckles because the controller has not completely removed them.

\subsection{Comparison with conventional EFC}

\begin{figure*}[ht]
\includegraphics[width=\textwidth]{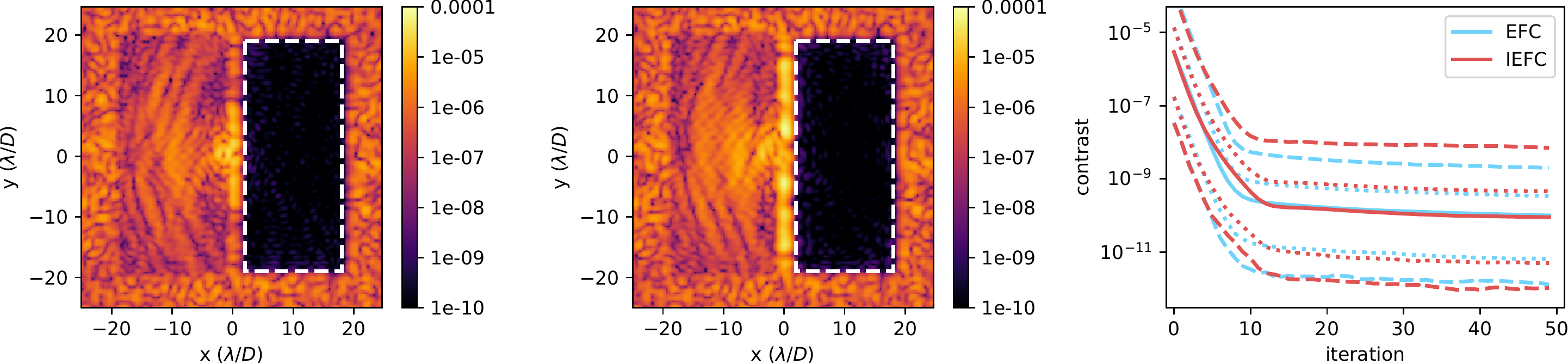}

\caption{\edited{A comparison between conventional EFC and iEFC. The left figure shows the post-coronagraphic image of EFC and the middle figure shows the post-coronagrahpic image of iEFC. The distribution of the contrast as a function of time is shown on the right. The blue and red line show the results for EFC and iEFC, respectively. The various line styles represent different quantiles; the dashed lines represents the 1\% and 99\% quantiles, the dotted lines follow the 16\% and 84\% quantiles and the solid line is the 50\% quantile (median contrast). Both methods converge within 10 iterations to the same median contrast of a few times $10^{-10}$. The 99\% quantile shows that iEFC has several brighter pixels than EFC.}}

\label{fig:comparison_efc}

\end{figure*}

\edited{The performance of iEFC is benchmarked against an implementation of conventional EFC. For both methods we use the exact same input disturbance (PSD with $\lambda/10$ peak-to-valley power and a -2.5 exponent), coronagraph (SCV) and dark hole region. The results are shown in Figure \ref{fig:comparison_efc}. The results shown equal performance between EFC and iEFC. The convergence rate and final contrast are similar. The only difference is that iEFC has brighter peak pixels, which is shown by the difference in the contrast levels of the 99th percentile. For iEFC the center region is very bright. This is mainly due to the selected single actuator probes which have low sensitivity at those areas \citep{potier2020comparing}. Conventional EFC regularizes the control of the modes from that region better. However, overall EFC and iEFC have the same performance.}

%A schematic of MagAO-X can be seen in Figure XXX. 
\section{Demonstration with MagAO-X}
\subsection{MagAO-X setup}
In this section, we verify the iEFC algorithm in the lab with MagAO-X. The optical layout of MagAO-X is split into two optical benches that are connected by a periscope. MagAO-X uses a woofer-tweeter architecture with an ALPAO-97 DM as woofer and a Boston Micromachines 2K tweeter \citep{males2018magaox, close2018magaox, males2022magao}. The system accepts an f/11 beam on the upper optical bench that first hits the woofer and then the tweeter. This beam is relayed to the lower bench where the pyramid wavefront sensor (PWFS) and the science instruments sit. The wavefront sensor beamsplitter splits the light into two paths, one for the PWFS and one for the science instruments. An additional ALPAO-97 DM was placed directly after the beamsplitter in the science path. This non-common path correction (NCPC) DM, is used for low-order NCPA control. After the NCPC DM the light passes through a preapodizer wheel. An off-axis parabola (OAP) is used to bring the beam to a focus in a focal plane, where a filter wheel is placed to switch between different focal plane masks. After another OAP, the beam is collimated and passes through the Lyot Stop filter wheel. The final OAP creates a F/69 beam onto the science cameras. The science cameras sample the PSF with 3 pixels per $\lambda/D$ at H$\alpha$.

\subsection{Low-order NCPA stabilization}
MagAO-X has a Low-Order Wavefront Sensor (LOWFS) to stabilize the PSF on the coronagraphic focal plane masks. All the currently installed focal plane masks are reflective. The reflection from the focal plane mask is reimaged onto the LOWFS camera. The reimaged PSF is used to stabilize low-order NCPA by feeding back a signal to the NCPC DM. The wavefront sensing is done with either  centroid tracking for tip-tilt control or linearized phase diversity for several low-order modes. Linear phase diversity has a small dynamic range, which for is sufficient for our purpose of low-order NCPA stabilization. The image is defocused as much as possible by translating the camera by 15 mm out of focus. Then an interaction matrix is built up using the calibration strategy from CACAO \citep{guyon2018compute,guyon2020adaptive}. In this calibration strategy, a set of Hadamard and Fourier modes are used to calibrate the response between the DM and the wavefront sensor. The controlled modes are limited to tip/tilt and focus for the experiments in this work. %\todo{Figure with PSD's and rejection transfer functions in an appendix?}

\subsection{MagAO-X dark hole results}
\begin{figure}[ht]
\includegraphics[width=\columnwidth]{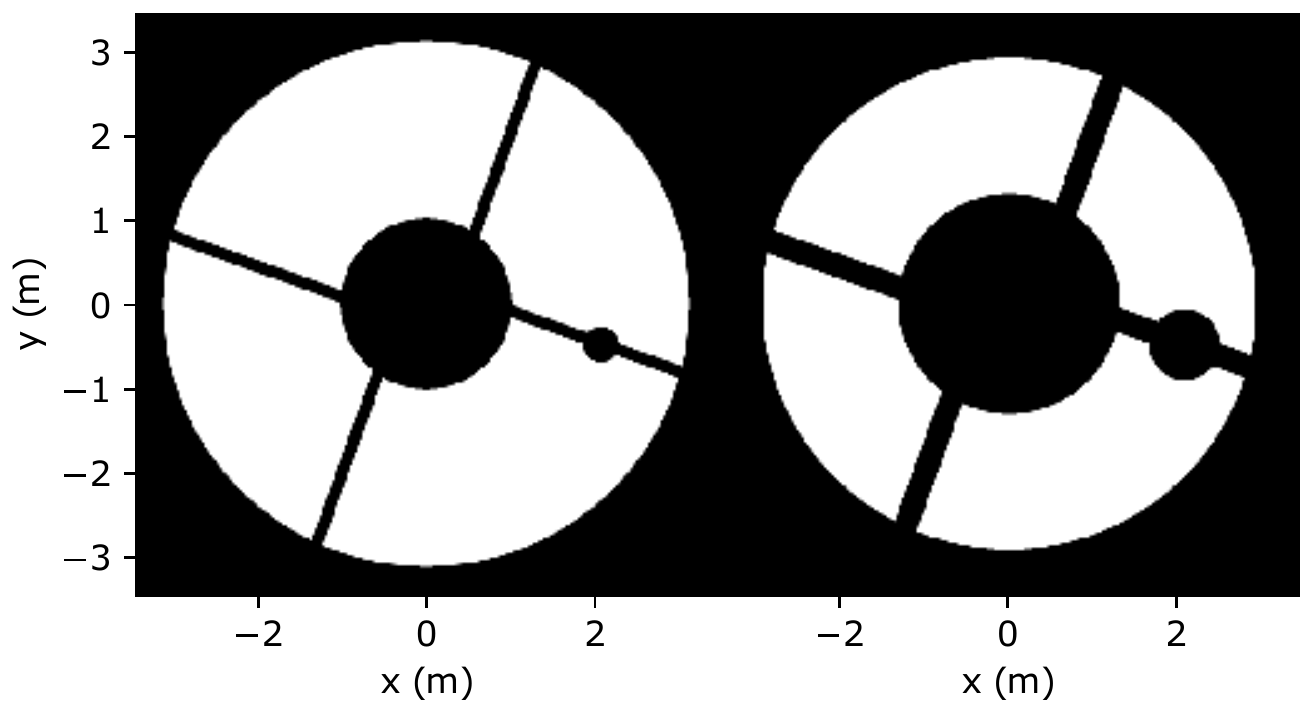}
\caption{Pupil apodizer (left) and Lyot stop (right) of MagAO-X for the coronagraphs. Both masks block an additional small circular patch on the right side of the pupil to mask a surface defect on the DM.}
\label{fig:magaox_pupil}
\end{figure}

There are currently 3 different coronagraphs installed in MagAO-X\footnote{see https://magao-x.org/docs/handbook/observers/coronagraphs.html for details of the MagAO-X coronagraphs.}, a classic Lyot Coronagraph (CLC), the vector Apodizing Phase Plate (vAPP) and the knife-edge Phase Apodized Lyot Coronagraph (PAPLC). We only apply iEFC on the CLC and the PAPLC. Both the CLC and PAPLC use the same entrance aperture and Lyot stop, which are shown in Figure \ref{fig:magaox_pupil}. The preapodizer of MagAO-X has a an additional masked area for a for a surface defect on the tweeter we call the "bump." The actuator itself is able to move but the surrounding actuators are not able to correct the bump. Therefore, the bump is masked to reduce its effect on the coronagraphic performance. The first created a one-sided dark-hole from 5 to 20 $\lambda/D$ with the CLC in a broadband filter (r-band). In the second experiment, we created a small inner-working angle one-sided dark-hole from 1.5 $\lambda/D$ to 15 $\lambda / D$ with the PAPLC in a narrowband filter (H$\alpha$). Most of the experiments have been done in the H$\alpha$ filter or r-band filter because those provided the highest photon flux in combination with the internal source.

\subsection{Classic Lyot Coronagraph}
\begin{figure}[ht]
\includegraphics[width=\columnwidth]{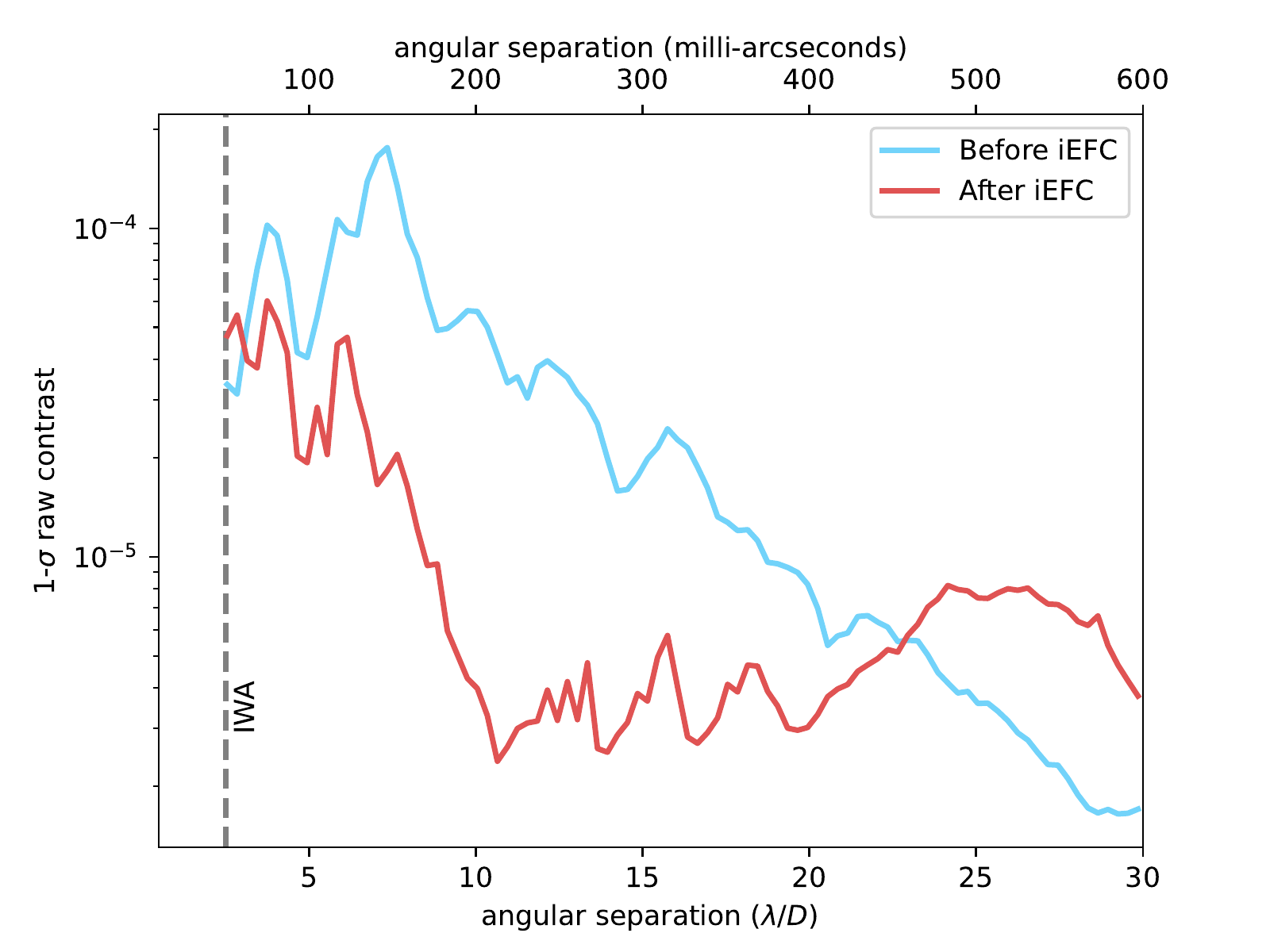}
\caption{Radial profile before and after iEFC with the CLC in the r-band filter. The blue line and red line show the contrast before and after wavefront control, respectively. The contrast is an order of magnitude lower after control.}
\label{fig:clc_radial_profile}
\end{figure}

\begin{figure*}[ht]
\includegraphics[width=\textwidth]{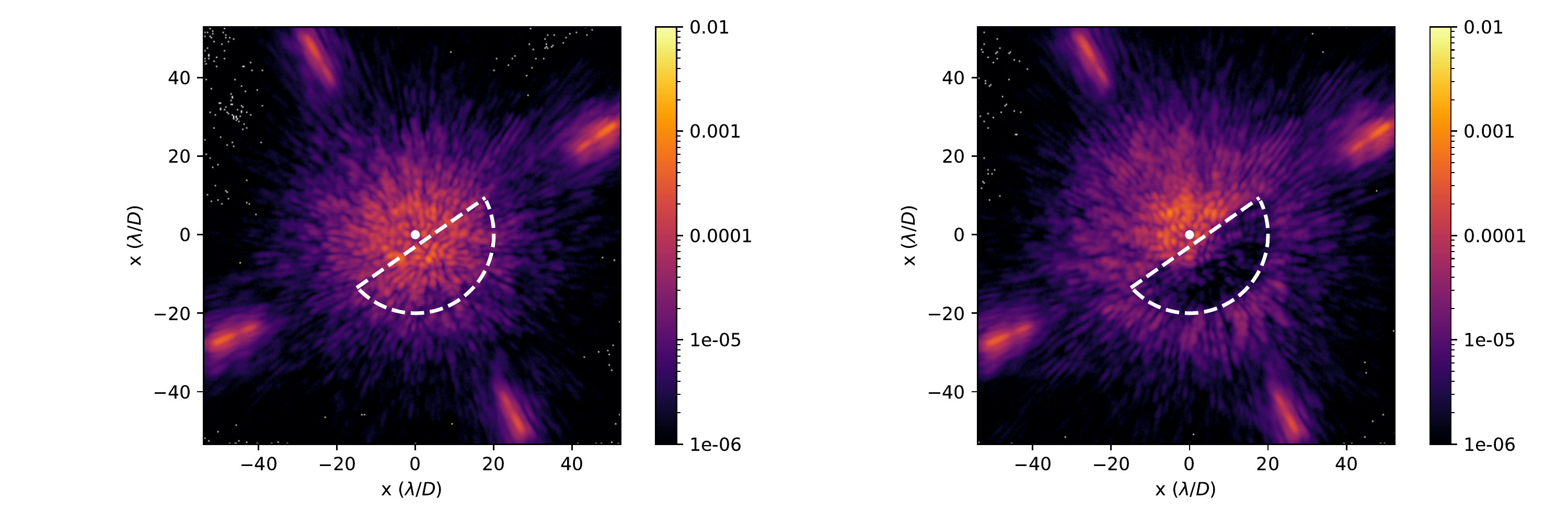}
\caption{D-shaped dark hole for the MagAO-X CLC. The inner-working angle is 3.0 $\lambda/D$ and the outer-working angle is 20 $\lambda/D$. The white border indicates the area that is nulled by iEFC. The center of the field is marked by the white dot. The left figure shows the coronagraphic focal plane before control and the right figure after control. Most of the dark hole after control is at a contrast between $10^{-6}$ and $10^{-5}$.}
\label{fig:bbclc_magaox}
\end{figure*}

The CLC uses a circular absorbing chrome mask. There are two different diameters available within MagAO-X. One with a radius of 3 $\lambda/D$ at H$\alpha$ and one with a radius of 5.0 $\lambda/D$ at H$\alpha$. The chrome dot reflects the light that hits it to the LOWFS. The optical density (OD) of the chrome dots depends on the wavelength with a OD of about 4.5 to 5.0 in r-band. We used the 3.0 $\lambda/D$ dot in r-band to create the dark hole. The radial profile before and after iEFC is shown in Figure \ref{fig:clc_radial_profile} and the full focal plane images before and after control are shown in Figure \ref{fig:bbclc_magaox}. \edited{This dark hole was created in 5 iterations of iEFC.}

The speckles in the D-shaped region are dimmed by about an order of magnitude with iEFC. The contrast gain is not as great as the simulations showed before. This is due to the chromatic aberrations inside the telescope simulator of MagAO-X. The telescope simulator is has a significant amount of chromatic defocus over r-band. Another source of chromaticity came from the Atmospheric Dispersion Compensator (ADC). The motors of our ADC prisms were not functional during this experiment. This left the ADC prisms in a fixed position that had residual dispersion. So not only was the PSF out of focus, it was also being dispersed by the ADC prisms. The combination of these two chromatic aberrations made it difficult to push to deep contrasts in broadband with the MagAO-X telescope simulator. However, iEFC was still able to reduce the speckle intensity to a couple times $10^{-6}$. Such contrast levels are enough to significantly improve ground-based observations even though the contrast is not that deep compared to space-based applications. The wind-driven halo will be the dominant factor after the intensity of the quasi-static speckles have been reduced to the level of $10^{-6}$ to $10^{-5}$ \citep{males2018lpc, cantalloube2020windhalo}.

\subsection{Knife Edge Phase Apodized Lyot Coronagraph}
The PAPLC uses a reflective knife edge focal plane mask that blocks all light that hits it \citep{por2020phase}. The knife edge is a silver coated Thorlabs D-shaped pickoff mirror. The mirror has a sharp edge and has a wedge removed so that the transmitted beam is not obstructed by the substrate. The knife edge was placed at the first null of the Airy pattern, which is at $1.1 \lambda / D$ for the Magellan pupil. The effective inner-working angle (IWA) is similar to the knife edge position. This particular configuration would create a small enough IWA to image an exoplanet like Proxima b \citep{anglada2016terrestrial} that has a angular separation of 45 mas at maximum elongation. The PAPLC is intrinsically one-sided so the dark hole will always be one-sided. The PAPLC dark hole was created to go as deep as possible with MagAO-X to determine the fundamental limit of the instrument. \edited{The shown dark hole was created after two rounds of 10 iterations of iEFC. The camera exposure settings had to be changed in between rounds to account for the large difference in dynamic range.}

\begin{figure}[ht]
\includegraphics[width=\columnwidth]{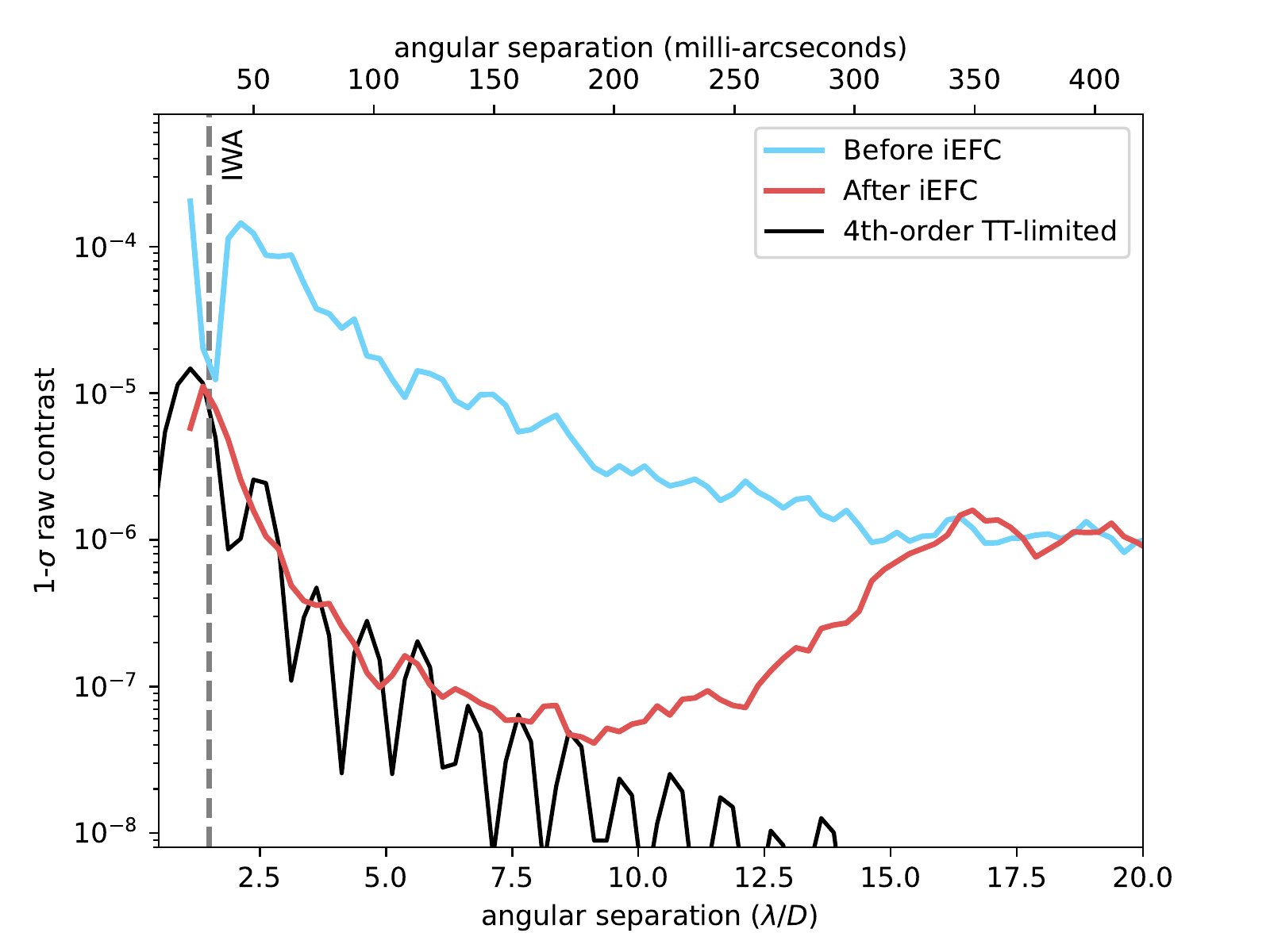}
\caption{Radial profile before and after iEFC with the PAPLC in the narrowband H$\alpha$ filter. The blue line and red line show the contrast before and after wavefront control, respectively. The contrast is two orders of magnitude lower after control. The black line shows the contrast profile for a theoretical 4th-order coronagraph that is limited by the measured MagAO-X jitter. The contrast after iEFC follows a similar decay as the 4th-order coronagraph.}
\label{fig:paplc_radial_profile}
\end{figure}

\begin{figure*}[ht]
\includegraphics[width=\textwidth]{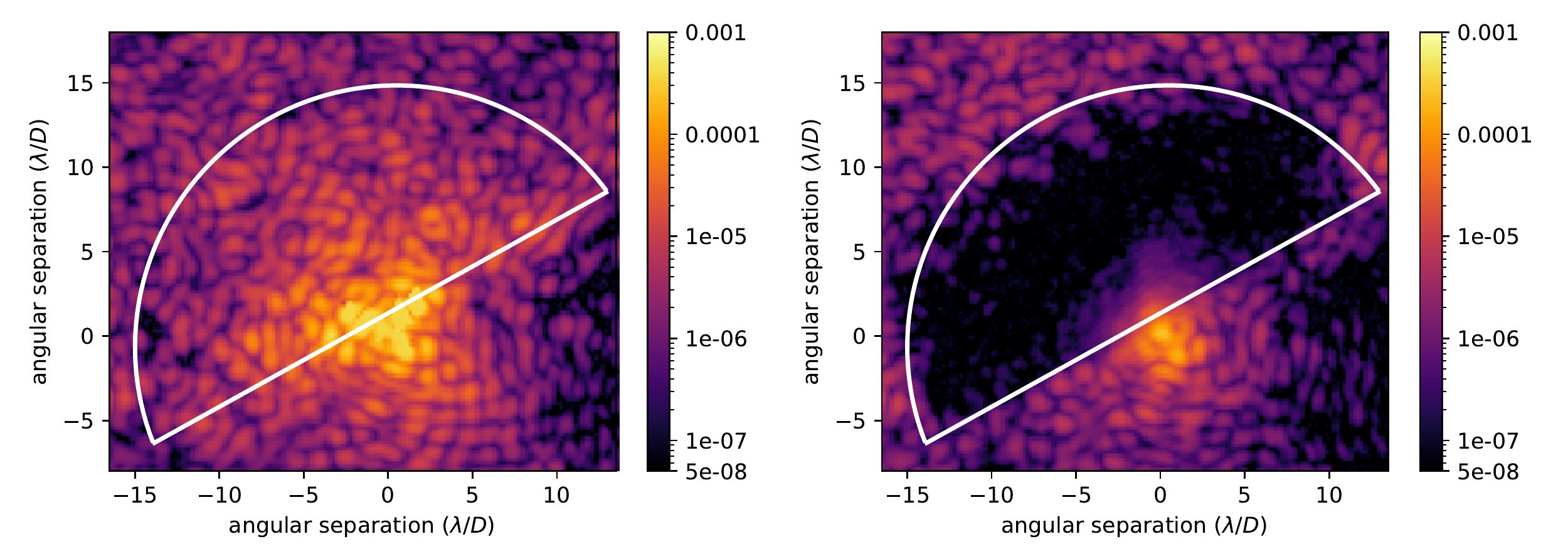}
\caption{A D-shaped dark hole for the MagAO-X PAPLC. The inner-working angle is 1.5 $\lambda/D$ and the outer-working angle is 15 $\lambda/D$. The white border indicates the area that is nulled by iEFC. The left figure shows the speckles before control and the right figure after control. The brightest speckles before control are saturated. Most of the dark hole after control is below a contrast of $10^{-7}$.}
\label{fig:paplc_magaox}
\end{figure*}

 The radial profile before and after control is shown in Figure \ref{fig:paplc_radial_profile}. The contrast is improved by a factor of 20 to 200 depending on the angular separation. The deepest part of the dark hole reaches a contrast of $5\cdot 10^{-8}$. The contrast degrades quite steeply at smaller IWA. The full focal plane images in Figure \ref{fig:paplc_magaox} reveal the source of the speckles. The halo at $<5\lambda/D$ is quite smooth and it is made by high frequency low-order aberrations. The most obvious culprit would be residual vibrations inside the instrument. The LOWFS is controlling tip/tilt at a loop speed of 200 Hz. The effective bandwidth of the control loop is 20-30 Hz. Any vibration that has a higher frequency will not be rejected by the loop. The residual jitter in MagAO-X after the low-order control loop is about $\lambda/30$ to $\lambda/20$ depending on the precise feedback gain in the LOWFS. Figure \ref{fig:paplc_radial_profile} shows the contrast for a 4th-order perfect coronagraph with the measured tip/tilt jitter of MagAO-X included. The contrast profile follows the measured profile almost exactly. This shows that jitter is limiting the contrast on the internal source.

\subsection{Dark hole digging with AO residuals}
The last experiment tests the performance of iEFC with residual AO speckles. The AO halo is made by running residual AO phase screens across the DM. The inital phase screens are made using the standard Las Campanas Observatory atmospheric layers \citep{prieto2010giant,males2018lpc}. A system with a 2-frame delay and integrator gain of 0.4 is used to create residual atmospheric layers. The system goes through 300 frames of residual turbulence during the integration of a single probed image. The simulated phase screens were generated at 1 kHz, which means that the effective integration time of iEFC is 0.3 s per image. This is then repeated for all the probe images. The total measurement time of the pair-wise probing is about 1.2 s.

\begin{figure*}[ht]
\includegraphics[width=\textwidth]{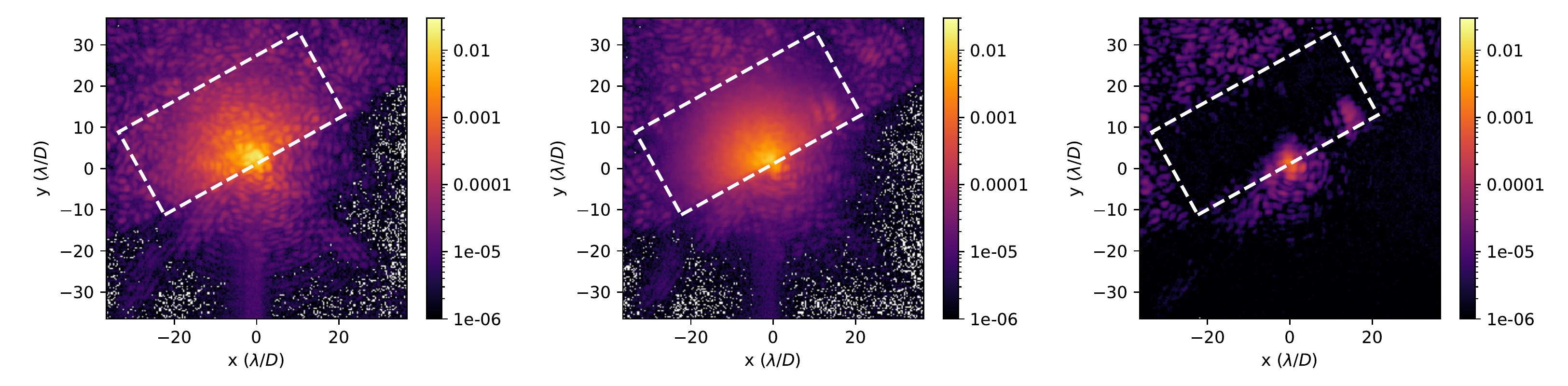}
\caption{Rectangular dark hole for the MagAO-X PAPLC. The inner-working angle is 1.5 $\lambda/D$ and the outer-working angle is 23 $\lambda/D$. The dark hole is 46 $\lambda/D$ wide. The white border indicates the area that is nulled by iEFC. The left figure shows the speckles with AO residuals before control and the middle figure after iEFC control. The right figure shows the same dark hole after iEFC but without AO residuals.}
\label{fig:aopaplc_magaox}
\end{figure*}

For this experiment, we created the largest possible dark hole with the PAPLC on MagAO-X. This dark hole had an inner-working angle of 1.5 - 2.0 $\lambda/D$ and an outer-working angle of 23 $\lambda/D$. The dark hole spans from -23$\lambda/D$ to 23 $\lambda/D$ in the other direction. The results are shown in Figure \ref{fig:aopaplc_magaox}. At the start the quasi-static speckles are stronger than the residual AO halo. However after 20 iterations all quasi-statics are not visible anymore and only the AO halo can be seen. There is one residual speckle near (10 $\lambda/D$, 10 $\lambda/D$). This speckle is caused by a orientation misalignment of the Lyot stop. Part of the spider leaks through and creates those speckles. This can also be seen in the lower left part of the image where there are two spider diffraction patterns with different orientations. The phase pattern that is made on the DM while running iEFC with turbulence is very similar to the phase pattern that is found without turbulence. The correlation coefficient between the two phase patterns is 0.905. This means that iEFC is converging to the same dark hole with and without AO residuals.

\section{Discussion and conclusion}
A new method for focal plane wavefront control has been described in this paper. The method makes use of the linear response between the DM commands and the differential images from pair-wise probing. The differential images are a measurable quantity which iEFC uses to measure an empirical calibration of the interaction matrix. This makes it much easier to implement \editedLC{because} a complex optical model of the system is not necessary anymore. This is a major advantage for ground-based systems that usually have more optics than their space-based counterpart. The ease of implementation is readily shown because iEFC has already been implemented on several other benches with different coronagraph architectures such as SCExAO \edited{that uses the Classic Lyot Coronagraph at H-band} \citep{ahn2021scexao,ahn2023efcldfc} and the Space Coronagraph Optical Bench (SCoOB) \citep{ashcraft2022space, van2022space} at Steward observatory \edited{which uses the knife-edge PAPLC and VVC}.

The simulations have shown that iEFC can achieve a deep contrast below $10^{-9}$ for many coronagraphs. The approach leads to similar performance levels as conventional EFC, making iEFC a competitive algorithm. \edited{The implementation of iEFC on MagAO-X showed a factor of 10 improvement in contrast for the CLC in r-band after only 6 iterations. The best contrast of $5\cdot10^{-8}$ at 7.5 $\lambda/D$ has been achieved with the PAPLC in a narrowband filter on MagAO-X. This required 20 iterations of iEFC.} The contrast level in the lab is not as deep as the contrast achieved in simulations. This is mainly attributed to the amount of tip-tilt jitter in MagAO-X. However, the achieved contrast is a much deeper than we expect to get on-sky because of the wind-driven halo created by the AO speckles \citep{males2018lpc, cantalloube2020windhalo}. The created dark holes will improve the performance of MagAO-X, if run on-sky, by at least an order of magnitude.

Pair-wise probing has been tricky to get implemented on-sky. One of the major challenges is offsetting the wavefront sensor so that it does not apply corrections for the probe and the dark-hole pattern. This is particularly challenging for systems that use pyramid wavefront sensors, which have a nonlinear response to offsets. So instead of using offsets, MagAO-X will use a dedicated DM in the coronagraph path for pair-wise probing and focal plane wavefront control. The new DM will be implemented as part of the MagAO-X phase II upgrade \citep{males2022magao}. This will separate the responsibilities of dark hole creation and AO correction to different control loops making it very easy to implement new focal plane wavefront control algorithms.

Another challenge for iEFC is the validity of the empirical calibration. We have not explored how robust the iEFC is against slow drifts in the system. The interaction matrix might have to be recalibrated every so often during the night. Broadband iEFC poses a problem too because a particular spectrum is assumed if the iEFC matrix is calibrated on the internal source. So any change in the source spectrum might change the performance of the algorithm. This could be solved by using an integral-field unit (IFU). Recently, an IFU was installed in MagAO-X (VIS-X)  \citep{haffert2022visible} that could be used for broadband speckle control. The conventional approach for EFC is to extract all the 2D images for all spectral channels. Then the electric field is measured with pair-wise probing for each channel. This approach not only depends on a good optical model of the instrument but also on a correct IFU extraction code. This can lead to many complications from model mismatches to spatial-spectral cross talk in the image reconstruction step. These issues can be completely circumvented by iEFC by directly minimizing the intensity of the spaxel spectra of the IFU because iEFC only needs access to intensity images. This makes iEFC a unique algorithm for IFU focal plane wavefront control. Applying iEFC to raw IFU images will be part of future work.

Focal plane wavefront control is necessary to reach the required contrast levels for Earth-like exoplanet imaging on the future large telescopes. However, few algorithms have been successful on-sky. Near future on-sky tests with iEFC on MagAO-X will be used to demonstrate its performance and viability.

\begin{acknowledgements}
We are very grateful for support from the NSF MRI Award \#1625441 (MagAO-X).

SYH was supported by NASA through the NASA Hubble Fellowship grant \#HST-HF2-51436.001-A awarded by the Space Telescope Science Institute, which is operated by the Association of Universities for Research in Astronomy, Incorporated, under NASA contract NAS5-26555.

KA acknowledges funding from the Heising-Simons Foundation.

The MagAO-X Phase II upgrade program is made possible by the generous support of the Heising-Simons Foundation.

\end{acknowledgements}

%-------------------------------------------------------------------

\bibliography{references}   % bibliography data in report.bib
\bibliographystyle{aa}   % makes bibtex use spiejour.bst

\end{document}